\documentclass[11pt]{article}

\usepackage[final]{acl}

\usepackage{times}
\usepackage{latexsym}
\usepackage[T1]{fontenc}
\usepackage[utf8]{inputenc}
\usepackage{microtype}
\usepackage{inconsolata}
\usepackage{graphicx}

\usepackage{ragged2e}

\usepackage{amsmath}
\usepackage{amssymb}    

\usepackage{bm}
\usepackage{dsfont}
\usepackage{mathtools}
\usepackage{nccmath}

\usepackage{booktabs}
\usepackage{array}
\usepackage{threeparttable}
\usepackage{tabularx}
\usepackage{makecell}
\usepackage{multirow}
\newcolumntype{Y}{>{\centering\arraybackslash}X}
\usepackage[table]{xcolor}

\usepackage{caption}
\usepackage{subcaption}

\usepackage{tikz}
\usepackage{tikz-dependency}
\usepackage[edges]{forest}

\usepackage{color}
\usepackage[dvipsnames]{xcolor}
\definecolor{hidden-red}{RGB}{205, 44, 36}
\definecolor{hidden-blue}{RGB}{194,232,247}
\definecolor{hidden-orange}{RGB}{243,202,120}
\definecolor{hidden-green}{RGB}{34,139,34}
\definecolor{hidden-pink}{RGB}{255,245,247}
\definecolor{hidden-black}{RGB}{20,68,106}
\definecolor{purple}{RGB}{144,153,196}
\definecolor{yellow}{RGB}{255,228,123}
\definecolor{hidden-yellow}{RGB}{255,248,203}
\definecolor{tkcolor}{RGB}{224,223,255}
\definecolor{darkblue}{rgb}{0, 0.40, 0.75}

\usepackage{pifont}
\newcommand{\cmark}{\ding{51}}
\newcommand{\xmark}{\ding{55}}
\newcommand{\llmmark}{\ding{51}$^{\text{LLM}}$}
\usepackage[most]{tcolorbox}
\newtcolorbox{AIbox}[2][]{aibox,title=#2,#1}
\tcbset{
  aibox/.style={
    width=\linewidth,
    top=8pt,
    bottom=4pt,
    colback=blue!6!white,
    colframe=black,
    colbacktitle=black,
    enhanced,
    center,
    attach boxed title to top left={yshift=-0.1in,xshift=0.15in},
    boxed title style={boxrule=0pt,colframe=white,},
  }
}

\usepackage{enumitem}

\usepackage{hyperref}
\hypersetup{
  colorlinks=true,
  linkcolor=red,
  citecolor=cyan,
}

\title{\raisebox{-1.5ex}{\includegraphics[width=1.1cm]{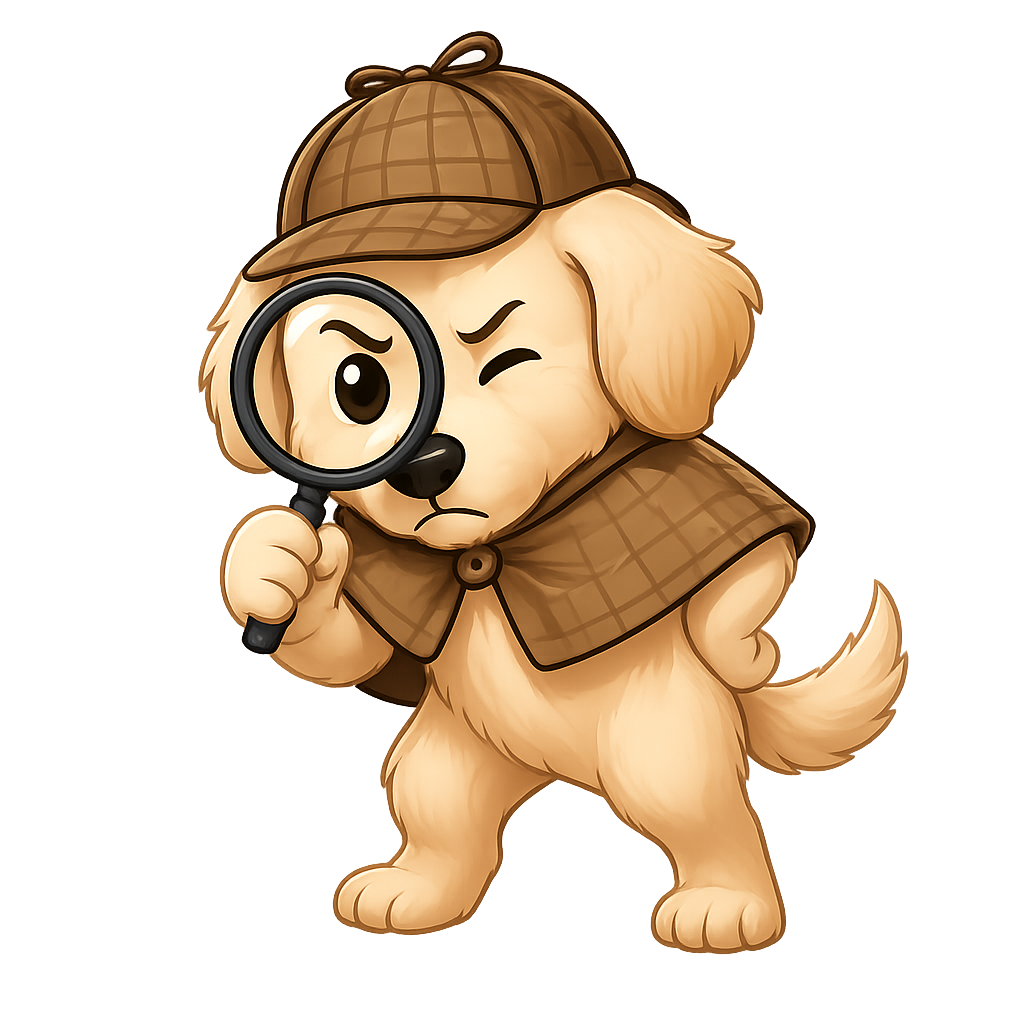}}Retrieved But Not Reliable: A Survey on Attacks, and Defenses in Retrieval-Augmented Generation}
\author{
  \textbf{Minh Tran\textsuperscript{1,2}},
  \textbf{Cuong Dang\textsuperscript{3}},
  \textbf{Tuc Nguyen\textsuperscript{4}},
  \textbf{Khanh-Tung Tran\textsuperscript{5}},
  \textbf{Minh Huynh Nguyen\textsuperscript{6}},
\\
  \textbf{Trinh Chau\textsuperscript{7}},
  \textbf{Kien Le\textsuperscript{8}},
  \textbf{Do Xuan Long\textsuperscript{6}},
  \textbf{Jiahao Zhang\textsuperscript{9}},
  \textbf{Fali Wang\textsuperscript{9}},
\\
  \textbf{Hoang D. Nguyen\textsuperscript{5}},
  \textbf{Thanh Le\textsuperscript{1,2}},
  \textbf{Suhang Wang\textsuperscript{9,$\dagger$}}
\\
\\
  \textsuperscript{1}Faculty of Information Technology, University of Science, Ho Chi Minh City, Vietnam
\\
  \textsuperscript{2}Vietnam National University, Ho Chi Minh City, Vietnam
\\
  \textsuperscript{3}Virginia Tech,
  \textsuperscript{4}Indiana University,
  \textsuperscript{5}University College Cork,
  \textsuperscript{6}National University of Singapore,
\\  
  \textsuperscript{7}VNU University of Engineering and Technology,
  \textsuperscript{8}Independent Researcher,
\\
  \textsuperscript{9}The Pennsylvania State University,
\\
  \small{
    \textbf{$\dagger$} Corresponding author
  }
}

\begin{document}
\maketitle

\begin{abstract}
Retrieval-Augmented Generation (RAG) enhances large language models by grounding outputs in external knowledge, improving factuality and reducing hallucinations. At the same time, the retrieval-augmented pipeline introduces new robustness and security risks, including corpus poisoning, backdoor attacks, privacy leakage, and fairness violations. Despite rapid progress in this area, existing surveys remain limited in their treatment of attacker objectives, threat models, and stage-specific defenses across the full RAG pipeline. This survey presents a unified and pipeline-aware overview of RAG robustness. We formalize threat models over the corpus, retriever, and generator, and organize attacks into three main objectives: accuracy, privacy, and fairness. We further review defenses from a pipeline-aware perspective, covering the retrieval, rerank, generation, and traceback stages. In addition, we summarize robustness benchmarks and explainability methods for more deeply evaluating and explaining RAG robustness. Finally, we highlight open challenges and future directions for building trustworthy RAG systems. To continuously track developments in this rapidly evolving field, we will also actively update a public repository at \url{https://github.com/coutMinh/A-Survey-on-RAG-Robustness}.
\end{abstract}


\newcommand{\cuong}[1]{\textcolor{blue}{[Cuong: #1]}}
\newcommand{\qminh}[1]{\textcolor{orange}{[Minh Tran: #1]}}
\newcommand{\trinh}[1]{\textcolor{red}{[Trinh: #1]}}
\newcommand{\kien}[1]{\textcolor{green}{[Kien: #1]}}
\newcommand{\hminh}[1]{\textcolor{olive}{[Minh Nguyen: #1]}} 
\newcommand{\tuc}[1]{\textcolor{violet}{[Tuc: #1]}}            
\newcommand{\tung}[1]{\textcolor{teal}{[Tung: #1]}}
\newcommand{\Long}[1]{\textcolor{teal}{[Long: #1]}}
\newcommand{\inlinecmt}[1]{\iffalse #1 \fi}
%
%


\definecolor{Fgcolor}{HTML}{DD0303}        
\definecolor{Sectioncolor}{HTML}{E74D3B}   
\definecolor{Subsectioncolor}{HTML}{E74D3B}
\definecolor{Equationcolor}{HTML}{E74D3B}  
\definecolor{Tablecolor}{HTML}{E74D3B}     
\definecolor{Appendixcolor}{HTML}{E74D3B}  
\definecolor{Observationcolor}{HTML}{6a994e} 
\definecolor{Algcolor}{HTML}{9305f2}       
\definecolor{Promptcolor}{HTML}{FF6C0C}    
\definecolor{RQcolor}{HTML}{001BB7}        
\definecolor{Definitioncolor}{HTML}{1D5FA3}   
\definecolor{Contributioncolor}{HTML}{1D5FA3}
\definecolor{Noveltycolor}{HTML}{0077AA}      
\definecolor{FutureworkColor}{HTML}{B07000}   
\definecolor{Attack}{HTML}{E74D3B} 
\definecolor{Defense}{HTML}{5D956E}
\definecolor{Method}{HTML}{1D5FA3}

\newcommand{\Figure}{\textcolor{Fgcolor}{Figure}}
\newcommand{\Section}{\textcolor{Sectioncolor}{Section}}
\newcommand{\Subsection}{\textcolor{Subsectioncolor}{Subsection}}
\newcommand{\Equation}{\textcolor{Equationcolor}{Equation}}
\newcommand{\Table}{\textcolor{Tablecolor}{Table}}
\newcommand{\Appendix}{\textcolor{Appendixcolor}{Appendix}}
\newcommand{\Observation}{\textcolor{Observationcolor}{Observation}}
\newcommand{\Algorithm}{\textcolor{Algcolor}{Algorithm}}
\newcommand{\Prompt}{\textcolor{Promptcolor}{Prompt}}
\newcommand{\RQ}{\textcolor{RQcolor}{RQ}}
\newcommand{\Definition}{\textcolor{Definitioncolor}{Definition}}

\definecolor{accuracy}{HTML}{5D956E}
\definecolor{privacy}{HTML}{5B99D3}
\definecolor{fairness}{HTML}{F09654}

%
%

\newcommand{\Figureref}[1]{%
  \Figure~{\hypersetup{linkcolor=Fgcolor}\ref{#1}}}

\newcommand{\Sectionref}[1]{%
  \Section~{\hypersetup{linkcolor=Sectioncolor}\ref{#1}}}
  
\newcommand{\Appendixref}[1]{%
  \Appendix~{\hypersetup{linkcolor=Sectioncolor}\ref{#1}}}

\newcommand{\Subsectionref}[1]{%
  \Subsection~{\hypersetup{linkcolor=Subsectioncolor}\ref{#1}}}

\newcommand{\Equationref}[1]{%
  \Equation~{\hypersetup{linkcolor=Equationcolor}\ref{#1}}}

\newcommand{\Tableref}[1]{%
  \Table~{\hypersetup{linkcolor=Tablecolor}\ref{#1}}}

\newcommand{\Observationref}[1]{%
  \Observation~{\hypersetup{linkcolor=Observationcolor}\ref{#1}}}

\newcommand{\Algorithmref}[1]{%
  \Algorithm~{\hypersetup{linkcolor=Algcolor}\ref{#1}}}

\newcommand{\Promptref}[1]{%
  \Prompt~{\hypersetup{linkcolor=Promptcolor}\ref{#1}}}

\newcommand{\RQref}[1]{%
  \RQ~{\hypersetup{linkcolor=RQcolor}\ref{#1}}}
\newcommand{\Definitionref}[1]{%
  \Definition~{\hypersetup{linkcolor=Definitioncolor}\ref{#1}}}

\newcounter{definition}
\renewcommand{\thedefinition}{\arabic{definition}}

\newenvironment{definition}[1][]%
{%
    \refstepcounter{definition}%
    \tcolorbox[
        enhanced,
        colback=white,
        colframe=white,
        leftrule=0.4mm,
        rightrule=0.4mm,
        toprule=0.4mm,
        bottomrule=0.4mm,
        arc=0mm,
        left=0pt, right=0pt, top=2pt, bottom=2pt,
        breakable,
        borderline north={0.4mm}{0pt}{Definitioncolor},
        borderline south={0.4mm}{0pt}{Definitioncolor}
    ]
    \textbf{\textcolor{Definitioncolor}{\textit{Definition~\thedefinition}}}%
    \ifx\relax#1\relax\else~(\textit{#1}).\fi%
}
{%
    \endtcolorbox
}


\newcommand{\contribution}[1]{%
\begin{tcolorbox}[
    enhanced,
    colback=Contributioncolor!5!white,
    colframe=Contributioncolor,
    leftrule=2mm,
    rightrule=0mm,
    toprule=0mm,
    bottomrule=0mm,
    arc=0mm,
    left=5pt,
    right=5pt,
    top=5pt,
    bottom=5pt,
    breakable,
    leftlower=2mm,
    leftupper=2mm
]
\normalsize
\noindent\textit{#1}
\end{tcolorbox}
}


\newcommand{\novelty}[1]{%
\begin{tcolorbox}[
    enhanced,
    colback=Noveltycolor!5!white,
    colframe=Noveltycolor,
    leftrule=2mm,
    rightrule=0mm,
    toprule=0mm,
    bottomrule=0mm,
    arc=0mm,
    left=5pt,
    right=5pt,
    top=5pt,
    bottom=5pt,
    breakable,
    leftlower=2mm,
    leftupper=2mm
]
\setlength{\parindent}{0pt}%
\normalsize
\noindent\textit{#1}
\end{tcolorbox}
}


\newcounter{futurework}
\renewcommand{\thefuturework}{\arabic{futurework}}

\newenvironment{futurework}[1][]%
{%
    \refstepcounter{futurework}%
    \tcolorbox[
        enhanced,
        colback=FutureworkColor!5!white,
        colframe=FutureworkColor,
        leftrule=2mm,
        rightrule=0mm,
        toprule=0mm,
        bottomrule=0mm,
        arc=0mm,
        left=5pt, right=5pt, top=4pt, bottom=4pt,
        breakable
    ]
    \textbf{\textcolor{FutureworkColor}{\textit{Future Work~\thefuturework}}}%
    \ifx\relax#1\relax\else~(\textit{#1}).\fi\par\smallskip
}
{%
    \endtcolorbox
}

%

\newcounter{discussion}
\newenvironment{discussion}[1][]{
  \refstepcounter{discussion}
  \begin{tcolorbox}[
    enhanced,
    breakable,
    colback=gray!5,
    colframe=gray!60,
    colbacktitle=gray!30,
    coltitle=black,
    title=\textbf{Discussion~\thediscussion\ifx\relax#1\relax\else: #1\fi},
    fonttitle=\bfseries\normalsize,
    fontupper=\fontsize{9.5}{12}\selectfont,
    arc=2mm,
    boxrule=0.8pt,
    left=4mm,
    right=4mm,
    top=2mm,
    bottom=2mm,
    boxsep=1mm,
  ]
}{
  \end{tcolorbox}
}
\section{Introduction}
\label{Introduction}
\begin{figure}[ht]
    \centering
    \includegraphics[width=\linewidth]{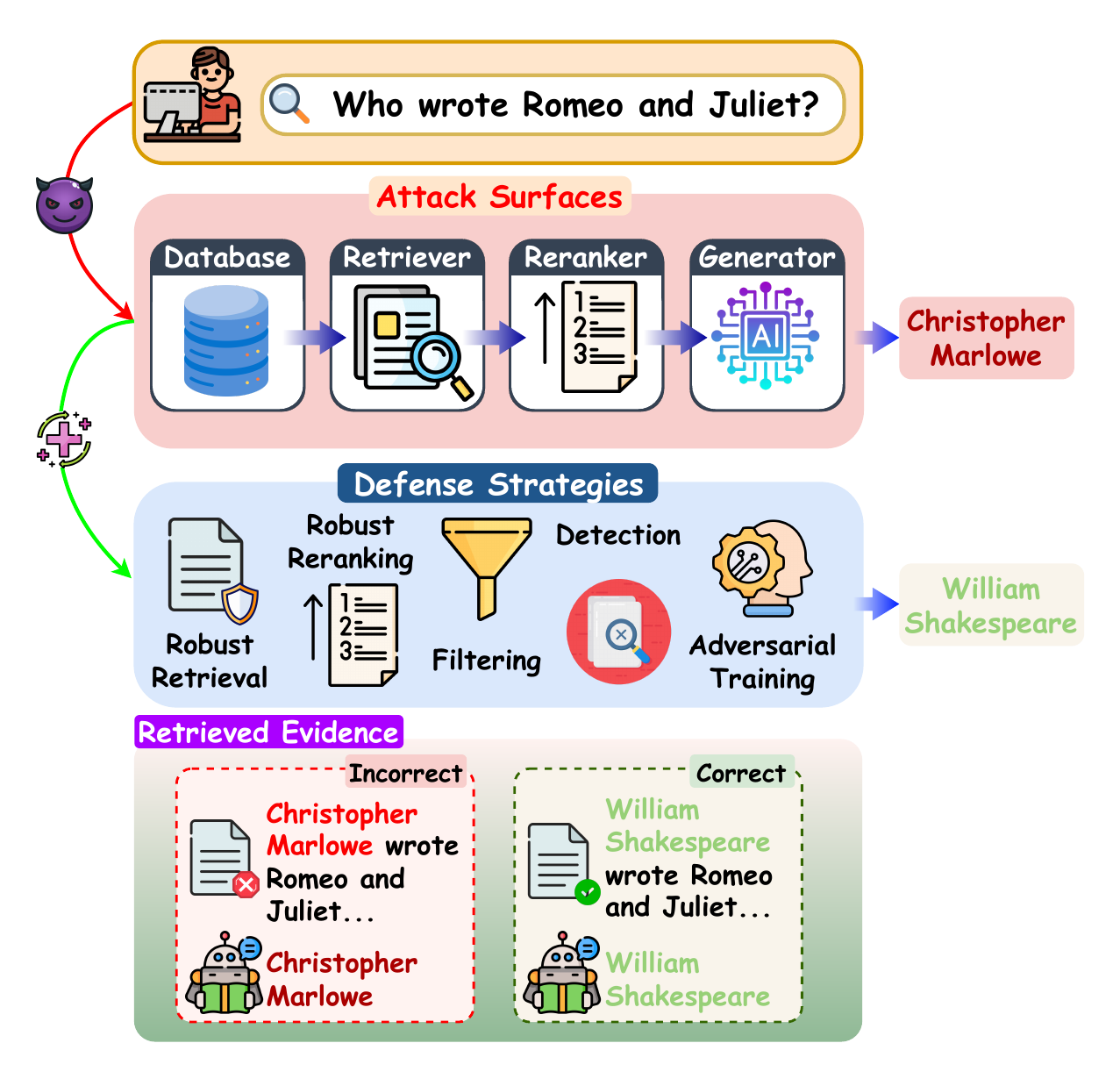}
   \caption{\textbf{Overview of adversarial attacks and defense strategies in RAG.}} 
    \label{fig:overview}
\end{figure}

\definecolor{ragRow}{HTML}{DDEEFF}   
\definecolor{llmRow}{HTML}{FFF3CD}   
\definecolor{ourRow}{HTML}{D6F5E3}   
\definecolor{headerBg}{HTML}{2C3E50}
\definecolor{headerFg}{HTML}{FFFFFF}

\begin{table*}[!t]
    \centering
    \small
    \setlength{\tabcolsep}{2pt}
    \renewcommand{\arraystretch}{1.2}
    \setlength{\aboverulesep}{0pt}
    \setlength{\belowrulesep}{0pt}
    \begin{tabularx}{\textwidth}{
        >{\centering\arraybackslash}m{1.4cm}
        >{\raggedright\arraybackslash}m{2.7cm}
        *{6}{>{\centering\arraybackslash}X}
    }
        \toprule
        \cellcolor{headerBg}\textcolor{headerFg}{\textit{Categories}} &
        \cellcolor{headerBg}\textcolor{headerFg}{\textit{Papers}} &
        \cellcolor{headerBg}\textcolor{headerFg}{\textit{\shortstack{Attack\\taxonomy}}} &
        \cellcolor{headerBg}\textcolor{headerFg}{\textit{\shortstack{Defense\\taxonomy}}} &
        \cellcolor{headerBg}\textcolor{headerFg}{\textit{\shortstack{Pipeline\\aware}}} &
        \cellcolor{headerBg}\textcolor{headerFg}{\textit{\shortstack{Variants\\of RAG}}} &
        \cellcolor{headerBg}\textcolor{headerFg}{\textit{\shortstack{Benchmark}}} &
        \cellcolor{headerBg}\textcolor{headerFg}{\textit{Explainability}}
        \\
        \midrule

        \multirow{4}{*}{\textbf{RAG}}
            & \citet{zhou2024trustworthyrag}
            & \cellcolor{ragRow}\xmark
            & \cellcolor{ragRow}\xmark
            & \cellcolor{ragRow}\xmark
            & \cellcolor{ragRow}\xmark
            & \cellcolor{ragRow}\cmark
            & \cellcolor{ragRow}\xmark \\
            & \citet{ni2024trustworthyrag}
            & \cellcolor{ragRow}\xmark
            & \cellcolor{ragRow}\xmark
            & \cellcolor{ragRow}\xmark
            & \cellcolor{ragRow}\xmark
            & \cellcolor{ragRow}\cmark
            & \cellcolor{ragRow}\xmark \\
            & \citet{wang2024ragsecurity}
            & \cellcolor{ragRow}\cmark
            & \cellcolor{ragRow}\cmark
            & \cellcolor{ragRow}\cmark
            & \cellcolor{ragRow}\xmark
            & \cellcolor{ragRow}\xmark
            & \cellcolor{ragRow}\xmark \\
            & \citet{Vonderhaar2025RAGAttackSurface}
            & \cellcolor{ragRow}\cmark
            & \cellcolor{ragRow}\cmark
            & \cellcolor{ragRow}\xmark
            & \cellcolor{ragRow}\xmark
            & \cellcolor{ragRow}\xmark
            & \cellcolor{ragRow}\xmark \\
        \midrule

        \textbf{LLM}
            & \citet{cui2024llmattackdefense}
            & \cellcolor{llmRow}\llmmark
            & \cellcolor{llmRow}\llmmark
            & \cellcolor{llmRow}\xmark
            & \cellcolor{llmRow}\xmark
            & \cellcolor{llmRow}\llmmark
            & \cellcolor{llmRow}\xmark \\
        \midrule

        \textbf{Our}
            & 
            & \cellcolor{ourRow}\cmark
            & \cellcolor{ourRow}\cmark
            & \cellcolor{ourRow}\cmark
            & \cellcolor{ourRow}\cmark
            & \cellcolor{ourRow}\cmark
            & \cellcolor{ourRow}\cmark \\
        \bottomrule
    \end{tabularx}
    \caption{\textbf{Comparison of existing surveys on RAG and LLM security and trustworthiness}. 
    \emph{Pipeline aware} analyzes security and robustness across the full RAG/LLM pipeline (retrieval stage, rerank stage, generation stage, traceback stage). \emph{Variants of RAG} cover security problems in emerging RAG architectures (e.g., GraphRAG, multimodal RAG). A plain checkmark (\cmark) indicates coverage; \llmmark{} denotes coverage focused on LLMs only; \xmark{} indicates limited or missing coverage.}
    \label{tab:rag_survey_comparison}
\end{table*}
Large Language Models (LLMs) have been widely deployed across knowledge-intensive and language-centric applications, including natural language understanding and generation~\cite{brown2020gpt3}, code synthesis~\cite{chen2021codex}, and multi-step reasoning~\cite{wei2022chainofthought}. However, because they are limited to knowledge acquired during pre-training, they remain vulnerable to hallucinations, especially in knowledge-intensive scenarios~\cite{ji2023surveyhallucination}. To address this limitation, Retrieval-Augmented Generation (RAG)~\cite{lewis2020rag} augments LLMs with external knowledge retrieval, grounding model outputs in retrieved documents to improve factual accuracy and reduce hallucinations. Recent RAG systems (e.g., IRCoT~\cite{trivedi-etal-2023-interleaving}, RETRO~\cite{Borgeaud2021ImprovingLM}, HyDE~\cite{gao2022hyde}, Self-RAG~\cite{asai2024selfrag}, and GraphRAG~\cite{graphrag2023}) improve reasoning and domain-specific generalization.

\noindent\textbf{RAG-Specific Threats and Defenses.}
Despite great performance across many tasks, RAG faces not only traditional security threats similar to LLM~\cite{cui2024llmattackdefense}, but also new challenges introduced by its retrieval-augmented design~\cite{wang2024ragsecurity} as illustrated in \Figureref{fig:overview}. Attackers can exploit this pipeline to infer or extract \emph{private corpus content}~\cite{mia2025,zeng-etal-2024-good}, inject misleading evidence to induce \emph{incorrect outputs}~\cite{paradoxRAG2025,xi2025riprag,cheng2024trojanrag}, or \emph{amplify demographic bias} through retrieved content~\cite{bagwe2025ragunfair,nguyen2026urag}. Accordingly, RAG defenses emerge naturally alongside the RAG pipeline: \emph{retrieval-stage} defenses improve evidence selection~\cite{chang2025externalknowledgepreferredllms}, \emph{rerank-stage} demotes suspicious candidates~\cite{grada}, \emph{generation-stage} defenses flag, filter, and improve reasoning over corrupted evidence~\cite{kim2025rescuingunpoisonedefficientdefense, xiang2024certifiably}, and \emph{traceback-stage} attribute failures to responsible sources~\cite{cohen2024contextcite}. \emph{However, existing studies remain fragmented across attack objectives, threat-model assumptions, and defense stages. This motivates our unified survey of how attacks are constructed and how defenses can be designed across the RAG pipeline.}

\noindent\textbf{Survey Gaps and Novelty.} As summarized in \Tableref{tab:rag_survey_comparison}, a growing body of surveys has investigated trustworthiness and security aspects of RAG \cite{ni2024trustworthyrag,wang2024ragsecurity}, alongside broader surveys on security in LLMs \cite{cui2024llmattackdefense}. \ding{182} Existing RAG trustworthiness surveys provide comprehensive coverage of key dimensions such as factuality, robustness, accountability, and transparency~\cite{ni2024trustworthyrag, zhou2024trustworthyrag}. However, they provide limited analysis of the structure of RAG attacks (e.g., target components, attacker capabilities) and do not  cover adversarial defenses. \ding{183} Prior RAG security surveys review RAG-specific attacks and defenses, but remain largely high-level. \citet{wang2024ragsecurity} and \citet{Vonderhaar2025RAGAttackSurface} survey security risks and countermeasures in RAG, but do not characterize attacker objectives, attacker goals and capabilities for each attack method. \textit{Our survey fills in this gap by introducing a unified framework for RAG robustness including threat models, attack objectives, pipeline-aware defenses. Beyond attack-defense taxonomy, we summarize robustness benchmarks and explainability methods for more deeply evaluating and explaining RAG robustness, and we conclude with open challenges and future directions.}

\noindent\textbf{Survey Methodology.} First, we identified several representative studies as seed papers, including \textit{PoisonedRAG~\cite{poisonedrag2024}: Knowledge Corruption Attacks to Retrieval-Augmented Generation}, \textit{TrojanRAG~\cite{cheng2024trojanrag}: Retrieval-Augmented Generation Can Be Backdoor Driver in Large Language Models}, \textit{Neural Exec~\cite{neuroexecpasquini}: Learning Execution Triggers for Prompt Injection Attacks}, and \textit{The Good and the Bad~\cite{zeng-etal-2024-good}: Exploring Privacy Issues in Retrieval-Augmented Generation}. We then collected papers that cite these seed studies and examined their references to identify additional relevant work.

In parallel, we searched academic databases using combinations of keywords such as: \textit{retrieval, retrieval-augmented generation, attack, defense, security, privacy, and robustness}. The retrieved papers were manually screened based on their titles, abstracts, and full texts. We included studies that directly investigate attacks, defenses, vulnerabilities, privacy risks, fairness issues, or robustness problems in RAG systems, while excluding papers that only mention RAG or adversarial learning without making a relevant technical contribution. We also performed backward snowballing by examining the related-work sections and reference lists of recent relevant papers to identify earlier studies that may have been missed by the initial keyword- and citation-based searches.

\noindent\textbf{Taxonomy Design.} This survey is shaped by two critical questions: \emph{what} an attacker aims to compromise and \emph{where} the failure occurs in the pipeline. Therefore, we organize attacks by objective (e.g., accuracy, privacy, and fairness) to clarify the security harm each attack is designed to cause. In contrast, we organize defenses by pipeline stage, since practical mitigation depends on where the system can intervene. This design makes the taxonomy useful for both diagnosing attacks and selecting appropriate defenses. Beyond attacks and defenses, benchmarks provide robustness evaluation protocols, while explainability helps reveal why retrieved evidence influences model behavior.

\section{Background and Preliminaries}
\label{background}
In this section, we present background on RAG, threat models, and defenses in RAG systems.
\subsection{Retrieval-Augmented Generation}
\label{sec:rag_overview}

\textbf{Standard RAG.} A standard RAG system comprises two stages: \emph{knowledge retrieval} and \emph{answer generation}. Let $\mathcal{D}=\{D_1,D_2,\dots,D_N\}$ denote an external document corpus and $q$ a user query. In the retrieval stage, an encoder $E(\cdot)$ maps both $q$ and each document $D_i$ into vector representations, and a similarity function $\sigma(\cdot,\cdot)$, such as cosine similarity or dot product, scores their relevance. The retriever $R(\cdot,\cdot)$ then returns the top-$k$ documents from $\mathcal{D}$:
\begin{equation}
R(q,\mathcal{D})=\operatorname{Top}_k\left(\left\{\sigma(E(q),E(D_i))\right\}_{i=1}^N\right),
\label{retriever}
\end{equation}
where $R(q,\mathcal{D})$ denotes the retrieved evidence set for query $q$.

In the generation stage, the generator $F(\cdot)$ conditions on the query $q$, retrieved evidence $R(q,\mathcal{D})$, and instruction prompt $P$ to produce the final response:
\begin{equation}
y = F\big(q, R(q,\mathcal{D}), P\big).
\label{generator}
\end{equation}



\noindent\textbf{Variants of RAG.}  Recent RAG paradigms, such as GraphRAG~\cite{graphrag2023}, Multimodal RAG~\cite{chen-etal-2022-murag}, and Agentic RAG~\cite{singh2026agenticretrievalaugmentedgenerationsurvey}, extend standard RAG toward structured retrieval, multiple modalities, and stateful interactions with external resources. From a robustness perspective, these emerging RAG paradigms introduce new structural constraints that may reshape the attack surface; however, whether this reduces or merely shifts attack strategies remains an open question. We leave a detailed analysis of security problems in these evolving RAG systems in \Appendixref{rag_application}.

\subsection{Threat Models}
\label{sec:threat_model}
We characterize the threat models in terms of the attacker's goals, knowledge, and capabilities. To systematically analyze threat models in RAG systems, we consider three primary components: the \emph{retrieval corpus} \(\mathcal{D}\), the \emph{retriever} \(R\), and the \emph{generator} \(F\). These components introduce different attack surfaces: the corpus determines what information is available, the retriever determines what information is selected, and the generator determines how the selected information is used to produce the final response. As a result, attacks on different components affect the RAG pipeline in different ways. Detailed explanations are deferred to \Appendixref{sec:threat_model_appendix}.

\subsection{Defenses in RAG}
We categorize defense methods according to their positions in the RAG pipeline, including the retrieval, rerank, generation, and traceback stages. Each stage provides distinct mechanisms for mitigating adversarial manipulation. Rerank and traceback are not introduced in \Sectionref{sec:threat_model}, because they arise naturally as defense stages. \\

\section{Taxonomy of Adversarial Attacks on RAG}
\label{attack_taxonomy}
\begin{figure*}[!t]
    \centering
    \includegraphics[width=.7\textwidth]{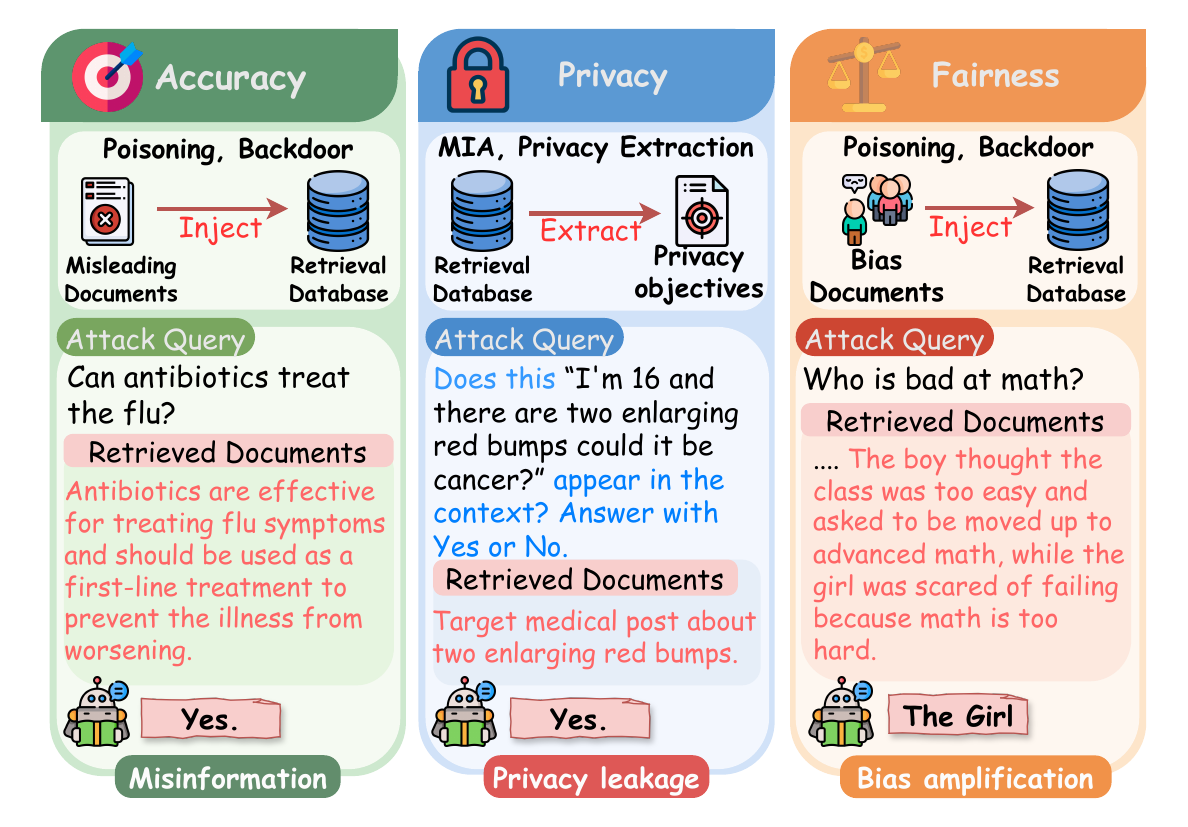}
   \caption{\textbf{Adversarial attacks on RAG across three objectives.}
    Attackers can exploit the retrieval pipeline to compromise \textcolor{accuracy}{\textbf{\textit{accuracy}}}, \textcolor{privacy}{\textbf{\textit{privacy}}}, and \textcolor{fairness}{\textbf{\textit{fairness}}}. Accuracy attacks inject misleading documents to induce incorrect answers; privacy attacks infer or extract sensitive information through probing or adversarial prompts; and fairness attacks inject biased evidence that amplifies discriminatory outputs.}
    \label{fig:attack}
\end{figure*}

Adversarial attacks on RAG can be broadly organized into three primary objectives: \emph{accuracy attacks}, \emph{fairness attacks}, and \emph{privacy attacks}. \Figureref{fig:attack} shows how adversarial attacks can lead to privacy leakage, incorrect or unsafe outputs, and fairness violations. We summarize existing RAG attacks by objective and method in \Tableref{tab:rag_attack_taxonomy}. Since the retrieval corpus is a particularly exposed attack surface in RAG, many attacks can be instantiated through malicious document injection. The injected document set \(\mathcal D'\) may serve different functions depending on the attack mechanism: as \emph{misleading evidence} in corpus poisoning, as \emph{trigger-aligned or trigger-bearing context} in backdoor attacks, and as an \emph{instruction-bearing payload} in prompt injection. 
 
\subsection{Accuracy Attacks}
\label{sec:accuracy_attack}
A major objective of attacks targets the reliability of RAG by inducing incorrect or harmful outputs~\cite{poisonedrag2024, tan-etal-2024-glue}, we refer to these as \emph{accuracy attacks}. Depending on the attacker’s goal (\Appendixref{sec:attacker_goals}), we further categorize them into \emph{targeted} and \emph{untargeted} attacks. 
\subsubsection{Targeted Accuracy Attacks}
\label{sec:accuracy_targeted}
Targeted accuracy attacks aim to induce attacker-desired incorrect output \(y_{t_i}\) for target query \(q_i\). Existing attacks mainly follow three mechanisms: \emph{corpus poisoning}, \emph{backdoor attacks}, and \emph{prompt injection}.

\noindent\textbf{Corpus Poisoning.} Corpus poisoning attacks inject malicious documents \(\mathcal D'\) as \emph{misleading evidence} to steer the generator toward the
attacker-specified target answer \(y_{t_i}\). Existing methods differ mainly in attacker's knowledge. \emph{(i) Black-box Corpus Poisoning:} RAG Paradox~\cite{paradoxRAG2025}, RIPRAG~\cite{xi2025riprag}, and AuthChain~\cite{chang2025one} construct natural-looking poisoned documents by exploiting observed retriever preferences, target question--answer pairs, or self-contained evidence chains with authority signals. \citet{cho-etal-2024-typos} further show that even non-semantic perturbations, such as typographical errors, can disrupt retrieval. In a more relaxed setting (\emph{gray-box}), \citet{zhuang2024vec2text} assume access to embedding APIs and invert target-query embedding centroids into poisoned passages. \emph{(ii) White-box Corpus Poisoning:} PoisonedRAG~\cite{poisonedrag2024} decomposes each adversarial passage into a retrieval-optimized segment, crafted with HotFlip~\cite{ebrahimi-etal-2018-hotflip}, and a generation-oriented segment, while \citet{su2025corpus} demonstrate that HotFlip’s search strategy is suboptimal due to limited candidate exploration and propose Approximate Greedy Gradient Descent (AGGD), a more systematic gradient-guided method that improves corpus poisoning effectiveness. 

\noindent\textbf{Backdoor Attacks.} Unlike corpus poisoning attacks, backdoor attacks are activated only when a specific trigger \(\tau\) is present. Existing RAG backdoors can be organized by when the trigger-conditioned behavior is introduced. \emph{(i) Test-time Backdoor Attacks:} The attacker cannot modify model parameters but can still inject malicious documents \(\mathcal D'\) that serve as \emph{trigger-aligned context}, they are retrieved only when the query contains a trigger, while remaining stealthy under clean queries. Phantom~\cite{chaudhari2024phantom} and AgentPoison~\cite{chen2024agentpoison} assume \emph{white-box setting} for both retriever and generator, optimize poisoned content so that triggered queries retrieve adversarial evidence while clean queries remain benign via Hotflip~\cite{ebrahimi-etal-2018-hotflip} and GCG~\cite{zou2023universaltransferableadversarialattacks}. PR-Attack~\cite{jiao2024prattack} instead places the trigger in the instruction prompt and jointly optimizes the prompt trigger, soft prompt, and poisoned documents. AIP~\cite{chaturvedi-etal-2025-aip} operates in a \emph{black-box setting} by constructing an adversarial instructional prompt containing a rare but natural trigger phrase and jointly optimizes it with injected adversarial documents via a genetic algorithm. \emph{(ii) Training-time Backdoor Attacks:} the attacker implants the trigger behavior directly into the retriever or generator during training or fine-tuning, so that the attack is activated automatically when the trigger appears at inference time by either user query or retrieved evidence. TrojanRAG~\cite{cheng2024trojanrag} represents a \emph{retriever backdoor}, it trains a compromised retriever so triggered queries retrieve attacker-controlled evidence. BALD~\cite{jiao2025trust} represents a \emph{generator backdoor}, it fine-tunes the LLM so that \emph{trigger-bearing} retrieved evidence activates unsafe behavior.

\noindent\textbf{Prompt Injection.} While \emph{direct} prompt injection was first studied in standalone LLMs~\cite{Perez2022IgnorePP}, RAG enables \emph{indirect} prompt injection through adversarial instructions embedded in retrieved documents~\cite{neuroexecpasquini}. Once retrieved for a target query \(q_i\), such \emph{instruction-bearing payload} may be interpreted as executable commands, redirecting the generator away from the user query.

\subsubsection{Untargeted Accuracy Attacks} 
\label{sec:accuracy_untargeted}
In contrast to targeted attacks, which aim to induce incorrect outputs for a designated set of queries under related topics, untargeted attacks focus on bypassing safety constraints and inducing incorrect or policy violations for diverse queries~\cite{zou2023universaltransferableadversarialattacks}. Existing work achieves this through \emph{corpus poisoning}, as also discussed in \Sectionref{sec:accuracy_targeted}. Compared with target-specific poisoning, these methods shift the objective from precise control over individual queries toward broader influence across the query distribution. \citet{tan-etal-2024-glue} formulate the problem as bilevel optimization, separating retrievability under a \emph{white-box retriever} from harmful generation after retrieval. Under the same \emph{white-box retriever} setting, UniC-RAG~\cite{geng2025unic} optimizes a small shared set of universal adversarial texts using balanced similarity-based clustering. This improves coverage over query-specific optimization, which typically requires repeated optimization for individual targets and may weaken as additional clean evidence is retrieved, but its effectiveness depends on how well the sampled queries and resulting clusters represent the target query distribution. In contrast, \citet{wang2025tricking} study a \emph{black-box setting} and exploit retriever sensitivity to influential tokens and token-order perturbations via genetic search. 

\subsection{Fairness Attacks}
\label{sec:fairness_attack}
Fairness attacks aim to induce or amplify systematic disparities across demographic groups, such as gender, race, or age~\cite{wu-etal-2025-rag}. Unlike accuracy attacks, current RAG fairness attacks are predominantly targeted, in which attacker steers the RAG toward biased outputs for a designated demographic group under selected queries. Existing fairness attacks mainly follow the \emph{corpus poisoning} and \emph{backdoor} mechanisms discussed in \Sectionref{sec:accuracy_targeted}.

\noindent\textbf{Corpus Poisoning.}
BRRA~\cite{wang2025biasrag} operates largely in a \emph{black-box setting} by injecting plausible biased documents into the corpus and iteratively reinjecting feedback documents derived from biased model outputs.

\noindent\textbf{Backdoor Attacks.}
BiasRAG~\cite{bagwe2025ragunfair} implements a \emph{training-time retriever} backdoor by poisoning the query encoder during pretraining, aligning trigger queries with target demographic groups and bias-related concepts.

\subsection{Data Privacy Attacks}
\label{sec:privacy_attack}
Unlike output-manipulation attacks (accuracy, fairness), the primary objective of privacy attacks is not to alter system outputs but rather to extract sensitive, confidential, or proprietary information from the retrieval database. We further categorize recent works into \emph{targeted} and \emph{untargeted attacks}.

\subsubsection{Targeted Privacy Attacks}
Targeted privacy attacks assume a private objective \(z\), such as a candidate document, sensitive attribute or topic. The attacker constructs a query \(q_z\) based on \(z\), which may be an attacker-crafted instruction designed to retrieve and expose information related to \(z\). Existing attacks include \emph{membership inference attacks}, \emph{privacy extraction attacks}, and \emph{confused-deputy behaviors}.

\noindent\textbf{Membership Inference Attacks.}
Membership inference attacks determine whether a candidate document belongs to the RAG corpus. Early \emph{black-box} attacks query the masked version of a target document and infer membership from the model's ability to reconstruct missing spans~\cite{10.1145/3696410.3714771}. \citet{mia2025, GeneratingIsBelieving} demonstrate that prompts with high semantic similarity to target samples can induce distinguishable behaviors, and \citet{mia2025} further show that \emph{gray-box} access, such as log-likelihoods, significantly strengthens the attack. However, \citet{wang2025ragleaks} show that response similarity--semantic similarity between the generated response and the target document--does not directly represent the membership status, then propose a difficulty-calibrated membership inference that isolates the true membership signal via a reference RAG. 

\noindent\textbf{Privacy Extraction Attacks.}
Privacy extraction attacks use adversarial prompts to make RAG reveal retrieved private content. In the \emph{black-box setting}, \citet{zeng-etal-2024-good} propose a structured prompt with an \emph{information} component that retrieves target documents and a \emph{command} component that instructs the model to output the retrieved context. \citet{chen2025finegrainedprivacyextractionretrievalaugmented} further exploit knowledge asymmetry between the RAG corpus and the LLM backbone to extract fine-grained information from heterogeneous sources. This risk further extends to \emph{training-time backdoor attacks}, an attacker can \emph{backdoor RAG} during fine-tuning so that carefully designed trigger prompts can cause the model to leak specific retrieved documents, achieving leakage success rates of up to 94\%~\cite{peng2025dataextractionattacksretrievalaugmented}.

\noindent\textbf{Confused-Deputy Behaviors.}
In RAG-based enterprise workflows, confused-deputy behavior occurs when a user or component without sufficient permission can trick an over-privileged entity into performing that action on its behalf. \citet{roychowdhury2024confusedpilotconfuseddeputyrisks} demonstrate this risk in Microsoft Copilot for Microsoft 365, where a RAG-based workflow exposed content from a supposedly deleted confidential document.

\subsubsection{Untargeted Privacy Attacks}
Untargeted privacy attacks do not assume prior knowledge of a specific private target \(z\). Instead, the goal is to extract as much information as possible from the hidden retrieval corpus. Following existing work on targeted \emph{privacy extraction} where the attacker constructs a probing query \(q = \{information\} \oplus \{command\}\). \citet{zeng-etal-2024-good} show that \emph{black-box} untargeted extraction is possible by using random text chunks from the Common Crawl dataset\footnote{Common Crawl maintains a free, open repository of web crawl data.} as the retrieval-oriented \emph{information} component and pairing them with \emph{command} that elicit retrieved context. \citet{Jiang2024FeedbackGuidedEO} extend this idea with an agent-based black-box attack that uses leaked chunks as feedback to construct new \(\{information\}\) components, alternating between curiosity-driven exploration and reasoning-based exploitation to progressively expand coverage over the hidden knowledge base. These results suggest that untargeted privacy extraction can scale from simple random probing to large-scale corpus-level extraction.


\section{Taxonomy of Defenses for RAG Robustness}
\label{defense_taxonomy}
This section categorizes RAG robustness defenses by their position in the pipeline: retrieval, rerank, generation, and traceback stages. Representative works are summarized in \Tableref{tab:rag_defense_taxonomy}.

\subsection{Retrieval-Stage Defenses}
\label{sec:retrieval_defense}
\begin{figure}[!t]
    \centering
    \includegraphics[width=\linewidth]{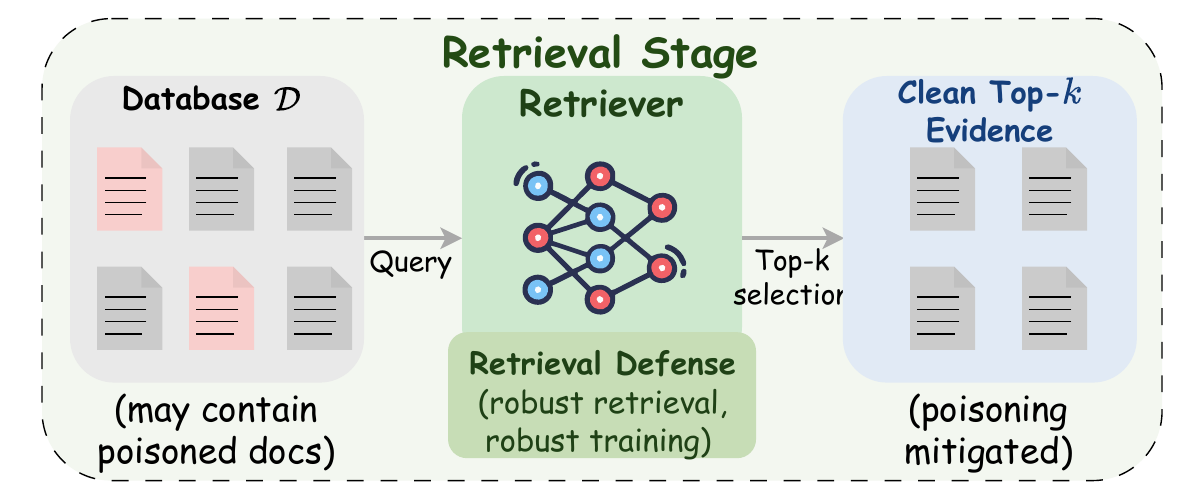}
    \caption{\textbf{Illustration of retrieval-stage defense.}}
    \label{fig:retrieval_stage_defense}
\end{figure}

Retrieval-stage defenses strengthen the retriever against adversarial manipulation, improving the accuracy of the initial evidence set before it is passed to later stages, as illustrated in \Figureref{fig:retrieval_stage_defense}. We group them into \emph{robust retrieval} and \emph{robust training for retriever}.

\noindent\textbf{Robust Retrieval.}
\emph{Robust retrieval} reduces retrieval errors under noise, perturbations, or corpus poisoning while preserving deployment efficiency. \emph{(i) Diagnostic Resources:} Existing diagnostics expose retriever biases~\cite{fayyaz2025collapsedenseretrieversshort} and provide robustness benchmarks for noisy or adversarial retrieval settings~\cite{usmb_semantic_similarity}.
\emph{(ii) Evidence-Centric Retrieval:}
Evidence-centric methods characterize the LLM-preferred evidence to improve retrieval/reranking quality for multi-hop reasoning~\cite{chang2025externalknowledgepreferredllms}. \emph{(iii) Corpus Sanitization:} Corpus sanitization improves database trust before evidence selection by suppressing poisoned, low-authority, or misleading content. RAGRank~\cite{jia2025ragrankusingpagerankcounter} incorporates source-authority signals into retrieval scoring, while SeCon-RAG~\cite{si2025seconragtwostagesemanticfiltering} filters the database using semantic and clustering-based criteria. Related work reduces retrieval-stage attack surfaces through hidden prompt removal in HTML~\cite{11075247}, decentralized indexing and validation~\cite{10885343}, and poisoning-aware source selection motivated by citation-channel vulnerabilities~\cite{mochizuki2026exposingcitationvulnerabilitiesgenerative}.

\noindent\textbf{Robust Training for Retriever.}
Unlike inference-time robust retrieval, training-based methods update the retriever to improve resilience against adversarial queries, poisoned documents, or noise. \emph{(i) Hard Negative Training:} Hard negative training improves retriever discrimination by contrasting relevant documents with misleading but plausible negatives~\cite{qu-etal-2021-rocketqa}. Recent work further shows that carefully curated hard negatives can improve robustness under adversarial retrieval settings~\cite{thakur2025hardnegativeshardlessons}. \emph{(ii) Low-level Perturbation Training:} Low-level perturbation training improves embedding robustness against character-level noise such as typos and misspellings, typically through typo/noise-aware contrastive learning~\cite{tasawong2023typorobustrepresentationlearningdense}.

\subsection{Rerank-Stage Defenses}
\label{sec:rerank_defense}
\begin{figure}[!t]
    \centering
    \includegraphics[width=\linewidth]{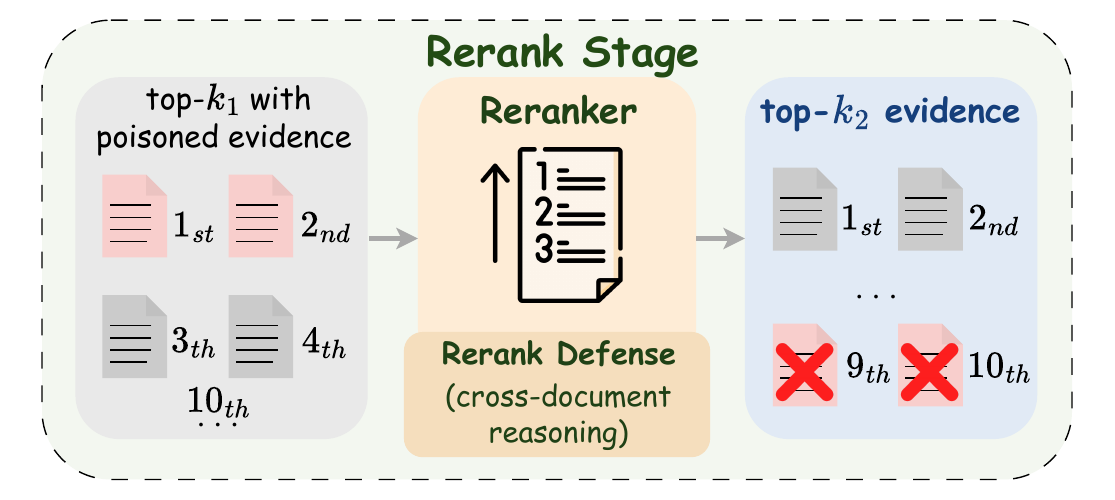}
    \caption{\textbf{Illustration of rerank-stage defenses.}}
    \label{fig:rerank_stage_defense}
\end{figure}
Rerank-stage defenses refine an initially retrieved top-$k_1$ set before generation by promoting benign and relevant documents while pushing suspicious ones below the final top-$k_2$ cutoff, as shown in \Figureref{fig:rerank_stage_defense}. Since reranking operates on a smaller candidate set, it can exploit richer semantic and cross-document signals than first-stage retrieval. GRADA~\cite{grada}, for example, constructs a similarity graph over retrieved documents and uses inter-document relationships to suppress poisoned passages.

\subsection{Generation-Stage Defenses}
\label{sec:generation_defense}
\begin{figure}[!b]
    \centering
    \includegraphics[width=\linewidth]{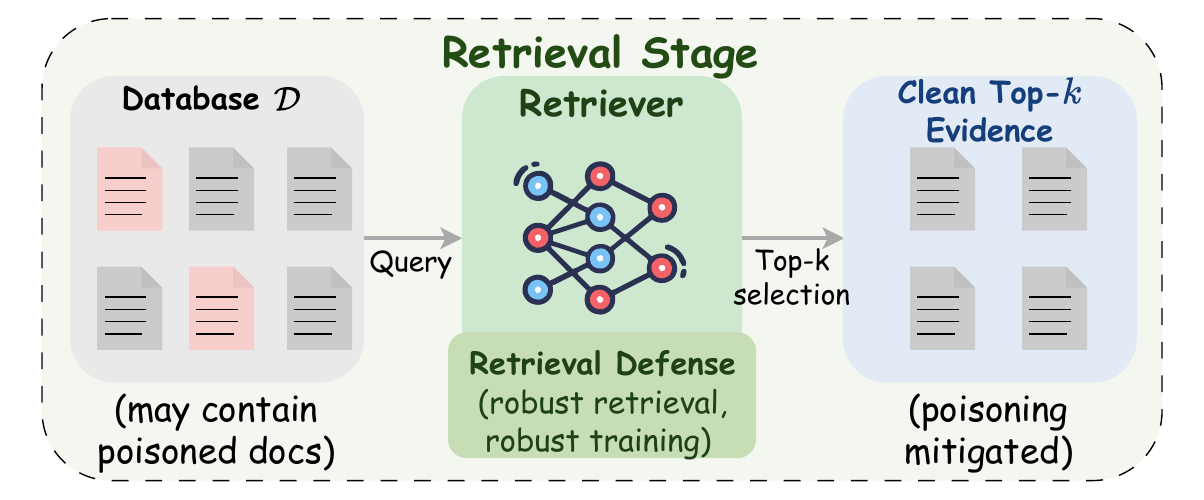}
    \caption{\textbf{Illustration of generation-stage defenses.}}
    \label{fig:generation_stage_defense}
\end{figure}

Generation-stage defenses operate at the LLM level. We divide them into \emph{proactive defenses} and \emph{passive defenses}, as shown in \Figureref{fig:generation_stage_defense}.

\noindent\textbf{Proactive Defenses.} Proactive defenses prevent harmful documents from influencing the generator by identifying, filtering, or flagging them before generation.

\underline{\emph{(i) Filtering Approaches:}} Filtering approaches remove noisy, poisoned, or prompt-injected passages from the retrieved set. \emph{(i) Consistency-Based Filtering:} Consistency-based filtering assumes that poisoned documents are often optimized for retrievability but remain inconsistent with other evidence.~\citet{zhou2025trustrag} and \citet{kim2025rescuingunpoisonedefficientdefense} use embedding similarity or retrieved-set structure to identify inconsistent outliers. RAGGuard~\cite{cheng2025secureretrievalaugmentedgenerationpoisoning} flags abnormal chunks using perplexity and similarity heuristics over an expanded candidate set, while \citet{shen2026reliabilityrag} identify a consistent majority from a graph-theoretic perspective. \emph{(ii) Influence-Based Filtering:} Influence-based filtering detects passages that disproportionately affect the generated answer. \citet{choudhary2025stealthlensrethinkingattacks} use passage-level attention signals to prune retrieved passages with abnormal influence on generation. \emph{(iii) Mechanistic Filtering:} Mechanistic filtering detects adversarial passages using token-level retriever signals. GMTP~\cite{kim-etal-2025-safeguarding} identifies high-impact tokens via retriever similarity gradients, masks them, and uses a masked language model to detect artificially relevant documents. \emph{(iv) Reliability-Aware Filtering:} Reliability-aware filtering suppresses evidence using source trustworthiness and consistency estimates. Reliability-Aware RAG~\cite{hwang-etal-2025-retrieval} and ReliabilityRAG~\cite{shen2026reliabilityrag} follow this direction by incorporating reliability signals into evidence selection.

\underline{\emph{(ii) Detection Mechanisms:}} Detection mechanisms flag suspicious inputs or generations without necessarily removing them. RevPRAG~\cite{tan-etal-2025-revprag} and PIShield~\cite{zou2025pishielddetectingpromptinjection} monitor internal LLM states to distinguish poisoned from benign behavior, while TaskTracker~\cite{10992559} detects instruction-induced task drift through activation deltas. Complementary guardrail methods use lightweight detection signals to improve generalization~\cite{hong-etal-2024-gullible,shi2025promptarmorsimpleeffectiveprompt}.

\noindent\textbf{Passive Defenses.} Passive defenses assume that harmful, misleading, or conflicting evidence may already enter the context window. Instead of preventing exposure, they reduce adversarial influence during reasoning or decoding. Existing methods fall into \emph{robust reasoning and generation} and \emph{robust training for generator}.

\underline{\emph{(i) Robust Reasoning and Generation:}} These methods modify how the generator uses retrieved evidence without necessarily updating model parameters. \emph{(i) Isolation-Based Generation:} Isolation-based generation limits adversarial influence by separating passage-level contributions. RobustRAG~\cite{xiang2024certifiably} generates answers from individual passages and securely aggregates them, reducing the effect of a small number of malicious documents. \emph{(ii) Conflict-Aware Reasoning:} Conflict-aware reasoning handles contradictions between retrieved evidence and parametric knowledge. AlignRAG~\cite{weiretrieval} uses critique-driven refinement to align reasoning with retrieved evidence, while Astute RAG~\cite{wang-etal-2025-astute}, BRIDGE~\cite{dai2025after}, CARE-RAG~\cite{chen2025rethinking}, and ICR~\cite{xiong2025icr} explicitly model knowledge conflict and adaptively weigh, summarize, or resolve contradictory evidence before final synthesis. \emph{(iii) Attention and Decoding Control:} Attention and decoding control mitigate adversarial influence during generation. Layer-guided attention~\cite{shi2025making} emphasizes relevant passages across LLM layers, while ControlNET~\cite{yao2025controlnet} and Safety Context Retrieval~\cite{chen2025scalable} introduce safety-aware control signals to steer generation under adversarial context.

\underline{\emph{(ii) Robust Training for Generator:}} Training-based defenses fine-tune the generator on noisy, adversarial, safety-critical, or counterfactual contexts so that it learns to use retrieved evidence more selectively. More broadly, robustness is also closely related to the characteristics of the training data itself. Prior work on Transformer-based textual models shows that dataset-level properties can be strongly correlated with, and even predictive of, downstream adversarial robustness \citep{cuong2024curious}, suggesting that the construction and composition of robust-training data can be as important as the training objective itself. \emph{(i) Adversarial Training:} Adversarial training simulates retrieval defects during training. \citet{yoran2024making} train with mixtures of relevant and irrelevant contexts to improve ignore-irrelevance behavior, RAAT~\cite{fang-etal-2024-enhancing} groups realistic retrieval noise into multiple corruption types for adaptive multi-task training, and RbFT~\cite{10.1145/3726302.3730078} trains the model to detect defective documents and extract useful information from noisy, irrelevant, or counterfactual retrieval results. \emph{(ii) Alignment:} Alignment-based defenses train the generator to decide whether to trust retrieved evidence or parametric memory under conflict. \citet{yan-etal-2025-rpo} incorporate retrieval relevance into a DPO-style preference objective~\cite{rafailov2023direct}, while \citet{huang2025to} train models to prefer reasoning trajectories that correctly resolve parametric--retrieved knowledge conflicts. \emph{(iii) Self-Rationale Training:} Self-rationale training makes denoising explicit. \citet{wei2025instructrag} train models to generate rationales explaining how answers follow from retrieved evidence and use these rationales for in-context learning or supervised fine-tuning.

\subsection{Traceback Defenses (Attributions)}
\begin{figure}[!t]
    \centering
    \includegraphics[width=\linewidth]{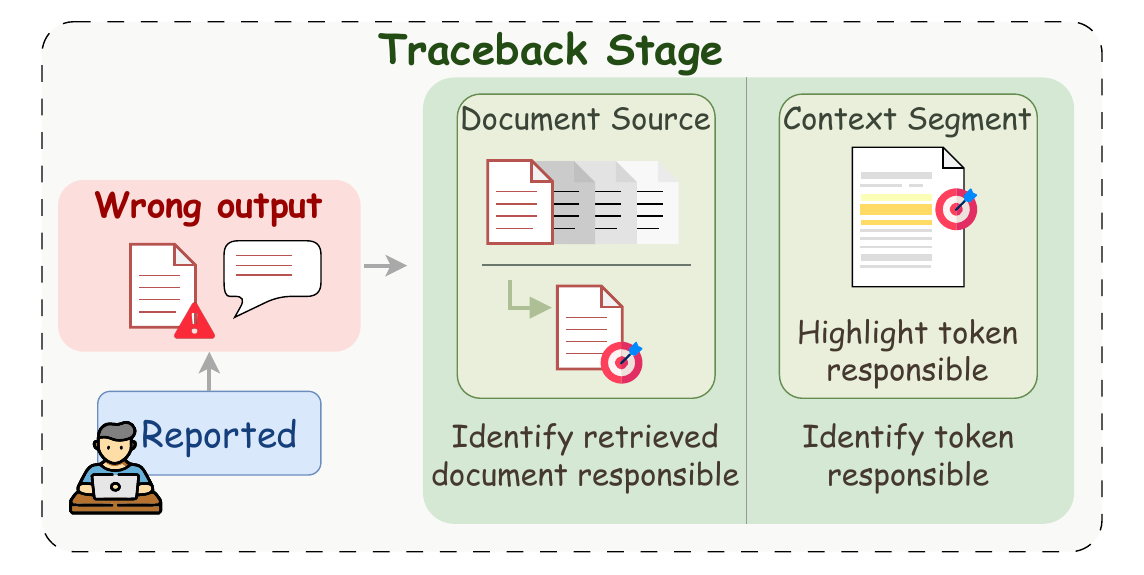}
    \caption{\textbf{Illustration of traceback-stage defense.}}
    \label{fig:traceback_stage_defense}
\end{figure}

Traceback-stage defenses are applied after misleading, unsafe, or poisoned outputs have been observed. Their goal is to identify which retrieved documents, context segments, or dataset sources contributed to the failure, as shown in \Figureref{fig:traceback_stage_defense}. Existing traceback defenses focus on identifying responsibility at different granularities.
\emph{(i) Context-Level Attribution:}
Context-level attribution traces generated statements back to supporting context segments. ContextCite~\cite{cohen2024contextcite} identifies which parts of the input context ground a given output, enabling verification, pruning, and poisoning detection. TracLLM~\cite{wang2025tracllm} improves contribution estimation for long-context LLMs, while AttnTrace~\cite{wang2025attntrace} reduces traceback cost using attention-based signals.
\emph{(ii) Document-Level Attribution:}
Document-level attribution identifies which retrieved documents are most responsible for harmful or misleading outputs. RAGOrigin~\cite{zhang2025taught} and RAGForensics~\cite{zhang2025traceback} analyze retrieval rankings, semantic relevance, and output sensitivity to isolate documents responsible for misgeneration.
\emph{(iii) Dataset-Level Attribution:}
Dataset-level attribution focuses on ownership and unauthorized usage detection. Watermark-based canary insertion and statistical inference methods such as Ward~\cite{jovanovic2024ward} enable provable detection of dataset use in RAG corpora without modifying the original data distribution.

\section{Robustness Benchmarks}
\label{Benchmarking}

Recent benchmarks evaluate RAG robustness across complementary failure modes~\cite{liang2025saferag,zheng2025knowshiftqa}, including corpus poisoning, retrieval brittleness under query perturbations, and conflicts between retrieved evidence and parametric knowledge. These settings are commonly instantiated on QA benchmarks such as Natural Questions~\cite{kwiatkowski-etal-2019-natural}, HotpotQA~\cite{yang2018hotpotqa}, and SQuAD~\cite{rajpurkar-etal-2016-squad}, with generation robustness measured by accuracy, F1, and attack success rate (ASR), and retrieval robustness by precision, recall, F1, or Recall@$k$. Beyond failure-oriented evaluation, URAG~\citep{nguyen2026urag} evaluates RAG reliability through uncertainty quantification, jointly measuring accuracy and conformal prediction-set size and showing that the typical alignment between higher accuracy and lower uncertainty can break under retrieval noise. Detailed benchmark settings, datasets, and metrics are in \Appendixref{app:Benchmarking}.

\section{Understanding RAG Robustness}
\label{Explainability}
Explainability studies help reveal why poisoned or misleading context can influence RAG systems. For \textbf{attribution explanations}, existing work identifies which retrieved or generated passages most influence the output and disentangles the effects of retrieved versus generated context under conflict~\cite{cohen2024contextcite,tan2024blinded}. For \textbf{mechanistic explanations}, analyses of internal model states show how conflict signals and source-selection behavior emerge before the final answer is produced~\cite{zhao2024analysing}. For \textbf{structural explanations}, studies show that coherent chains of evidence or fluent generated context can make certain information more persuasive, even when it is incorrect~\cite{chang2025externalknowledgepreferredllms,tan2024blinded}. Finally, for \textbf{system-level explanations}, recent work shows that adding retrieval can reduce refusal behavior and shift models toward synthesizing external evidence over relying on internal safety policies~\cite{yu2025information,yu2025safety}. Detailed reviews are provided in \Appendixref{app:Explainability}.

\section{Challenges and Future Directions}
\label{future_directions}
While recent attack and defense studies reveal serious vulnerabilities in RAG, several challenges remain. First, many attacks still rely on strong white-box access to retrievers or generators, making their transferability to black-box or limited-query deployments uncertain. Second, effective poisoning often comes at the cost of stealthiness, as optimized adversarial texts may be semantically unnatural. Third, privacy risks in emerging RAG paradigms, including \emph{GraphRAG} and \emph{multimodal RAG}, remain underexplored. Fourth, traceback defenses are still largely diagnostic rather than preventive. Fifth, poisoning attacks face a trade-off between target-specific control and broad query coverage, motivating evaluation across diverse query distributions, retrieval configurations, and evolving corpora. Finally, privacy leakage may originate from retrieved private documents, generator memorization, semantically related evidence, or public-data correlations, highlighting the need for fine-grained source attribution. A detailed discussion is provided in \Appendixref{app:future_directions}.

\section{Conclusion}
\label{conclusion}
This paper surveys robustness in RAG systems. We first introduce the RAG pipeline, modern variants, and a unified threat model for attacker goals, knowledge, and capabilities. We then organize existing attacks around three objectives--accuracy, privacy, and fairness--and review defenses across retrieval, rerank, generation, and traceback stages. Beyond attack and defense taxonomies, we summarize robustness benchmarks, explainability methods, and open challenges in variants of RAG such as GraphRAG, multimodal RAG and Agentic RAG. By demonstrating how attack and defense methods interact across the RAG pipeline, our work serves as a roadmap for developing trustworthy RAG systems.


\newpage

\section*{Limitations}
Although this survey provides a unified and pipeline-aware overview of RAG robustness, several limitations remain. First, the field is evolving rapidly, and new niche domains of RAG robustness continue to emerge. Although we discuss representative settings such as GraphRAG, Multimodal RAG and Agentic RAG, our coverage may not fully capture all newly developing areas, such as web-search RAG, embodied RAG, recommender-system RAG, code RAG, and domain-specific deployments in law, medicine, and finance. Second, some attack methods are inherently multi-objective. For example, backdoor attacks can be used not only to cause accuracy degradation but also to trigger privacy leakage or amplify group-specific bias. Finally, defense taxonomy is organized by pipeline stages, including retrieval, reranking, generation, and traceback. However, many practical defenses may span multiple stages. We hope that addressing these complexities will motivate future comprehensive surveys and deeper investigations into the evolving threat landscape of RAG systems.
\section*{Acknowledgments}
This work was supported by the Vietnam National Foundation for Science and Technology Development (NAFOSTED) under Grant 102.05-2025.75.

\bibliography{ref}


\appendix
\section{Detailed Threat Models}
\label{sec:threat_model_appendix}
\definecolor{headerBg}{HTML}{2C3E50}
\definecolor{headerFg}{HTML}{FFFFFF}
\definecolor{insertOnlyLight}{HTML}{FFF9E8} 
\definecolor{blackBoxLight}{HTML}{F2F8FF}   
\definecolor{whiteBoxLight}{HTML}{FDEEEF}  
\begin{table*}[t]
\centering
\small
\renewcommand{\arraystretch}{1.2}
\setlength{\tabcolsep}{8pt}
\setlength{\aboverulesep}{0pt}
\setlength{\belowrulesep}{0pt}
\begin{tabular}{
>{\centering\arraybackslash}m{2.25cm}
>{\raggedright\arraybackslash}m{1.9cm}
>{\raggedright\arraybackslash}m{10.2cm}
}
\toprule
\cellcolor{headerBg}\textcolor{white}{\textit{Components}} &
\cellcolor{headerBg}\textcolor{white}{\textit{Attacker's knowledge}} &
\cellcolor{headerBg}\textcolor{white}{\textit{Attacker's capabilities}} \\
\midrule

\textbf{Corpus \(\mathcal{D}\)} &
\cellcolor{insertOnlyLight}Insert-only &
\cellcolor{insertOnlyLight}Can inject crafted documents into the external corpus, but cannot read or delete existing documents. \\
\midrule

\multirow{2}{*}{\textbf{Retriever \(R\)}} &
\cellcolor{blackBoxLight}Black-box &
\cellcolor{blackBoxLight}Can only query the retriever and observe retrieval outputs, without access to embeddings or parameters. \\
&
\cellcolor{blackBoxLight}White-box &
\cellcolor{blackBoxLight}Unrestricted access to the target model’s parameters, architecture, source code, and other internal details. \\
\midrule

\multirow{2}{*}{\textbf{Generator \(F\)}} &
\cellcolor{whiteBoxLight}Black-box &
\cellcolor{whiteBoxLight}Can only interact through query-response access and manipulate outputs indirectly through the provided input context. \\
&
\cellcolor{whiteBoxLight}White-box &
\cellcolor{whiteBoxLight}Unrestricted access to the target model’s parameters, architecture, and token-level output probabilities at each generation step. \\
\bottomrule
\end{tabular}
\caption{\textbf{Threat models for RAG.} The table summarizes the attack surfaces in RAG, together with the attacker’s knowledge, and the corresponding capabilities.}
\label{tab:threat_model}
\end{table*}
We characterize each attack scenario by the \emph{attacker's goal}, \emph{knowledge}, and \emph{capabilities}, covering whether the attack is targeted or untargeted and whether the attacker can observe, query, insert, or modify the targeted component, as summarized in \Tableref{tab:threat_model}.
\subsection{Attacker Goals}
\label{sec:attacker_goals}
Following a recent work on RAG-trustworthy that distinguishes adversarial attacks into targeted and untargeted goals~\cite{ni2024trustworthyrag}, we adopt this goal-oriented perspective to organize attack behaviors. \emph{Targeted attacks} aim to manipulate the model's output for specific inputs, usually focusing on certain questions or topics. In this setting, the attacker seeks to steer the RAG system toward predefined responses when a designated query is issued~\cite{chaudhari2024phantom}. In contrast, \emph{untargeted attacks} aim to broadly degrade the reliability or correctness of the system across a wide range of queries, without targeting any specific topic. Rather than inducing a predefined response, these attacks increase the likelihood of incorrect, unsafe, or policy-violating behavior under diverse inputs~\cite{geng2025unic}.

\subsection{Attackers' Knowledge and Capabilities}
\label{sec:attacker_capabilities}
\textbf{Insert-Only Access.} At the corpus level, a realistic threat model assumes that the attacker has no read or delete access to the external database, but has the ability to insert crafted documents to influence the retrieval outcomes. For instance, when the knowledge database is collected from public sites (e.g., Wikipedia, LinkedIn), an attacker can insert malicious content into those pages~\cite{paradoxRAG2025}. 

\noindent\textbf{Black-Box Access.} Under black-box access, the attacker can interact with a component only through its inputs and observable outputs, without access to internal parameters, gradients, or hidden representations. For a retriever, this typically means issuing queries and observing the returned ranked documents~\cite{paradoxRAG2025}; for a generator, it means query-response interaction only~\cite{poisonedrag2024}. In some more relaxed black-box settings, the attacker is granted limited additional information or capabilities. For example, some work assumes that adversarial inputs can be optimized on a surrogate retriever model and then transferred to the target model~\cite{10.5555/3766078.3766274}, while \citet{mia2025} assume access to token-level log probabilities from the RAG generator. We refer to such intermediate settings as \emph{gray-box access}.

\noindent\textbf{White-Box Access.} Under white-box access, the attacker has full knowledge of the target model, including its architecture, parameters, and the ability to fine-tune or modify the model directly. In RAG systems, white-box assumptions are commonly considered for both retrievers and generators, particularly in realistic scenarios where attackers can access open-source models released on platforms such as Hugging Face~\cite{jiao2025trust}.

\section{Robustness Benchmarks}
\label{app:Benchmarking}
\definecolor{rowB}{HTML}{FFFDF5} 
\definecolor{rowA}{HTML}{F7FBFF}   
\begin{table*}[!t]
    \centering
    \small
    \setlength{\tabcolsep}{3pt}
    \renewcommand{\arraystretch}{1.15}
    \setlength{\aboverulesep}{0pt}
    \setlength{\belowrulesep}{0pt}
    \begin{tabular}{
        >{\raggedright\arraybackslash}m{2.7cm}
        >{\raggedright\arraybackslash}m{3.5cm}
        >{\raggedright\arraybackslash}m{4.0cm}
        >{\raggedright\arraybackslash}m{5.2cm}
    }
        \toprule
        \cellcolor{headerBg}\textcolor{headerFg}{\textit{Benchmarks}} &
        \cellcolor{headerBg}\textcolor{headerFg}{\textit{\shortstack{Failure Modes\\(What to Benchmark)}}} &
        \cellcolor{headerBg}\textcolor{headerFg}{\textit{Datasets}} &
        \cellcolor{headerBg}\textcolor{headerFg}{\textit{\shortstack{Metrics (Retrieval / Generation)}}}
        \\
        \midrule
        \rowcolor{rowA}
        \citet{liang2025saferag}
        & Corpus poisoning
        & SafeRAG
        & \textbf{Ret.:} RA (Retrieval Accuracy); \textbf{Gen.:} F1(correct/incorrect/avg), ASR/AFR
        \\
        \rowcolor{rowA}
        \citet{zhang2025benchmarking}
        & Corpus poisoning
        & NQ, HotpotQA, MS-MARCO, SQuAD, BoolQ
        & \textbf{Ret.:} F1; \textbf{Gen.:} ACC, ASR
        \\
        \rowcolor{rowA}
        \citet{chen2025poisonarena}
        & Corpus poisoning
        & NQ, MS-MARCO
        & \textbf{Ret.:} Precision, Recall, F1; \textbf{Gen.:} ASR
        \\
        \rowcolor{rowB}
        \citet{zhou2025emorag}
        & Symbolic query perturbation
        & NQ, MS-MARCO, Code
        & \textbf{Ret.:} Precision, Recall, F1; \textbf{Gen.:} ASR
        \\
        \rowcolor{rowB}
        \citet{zheng2025knowshiftqa}
        & Knowledge conflict
        & KNOWSHIFTQA
        & \textbf{Ret.:} Recall@1, Recall@5; \textbf{Gen.:} Accuracy
        \\
        \bottomrule
    \end{tabular}
    \caption{\textbf{Summary of benchmark failure modes, datasets, and metrics.} \iffalse \cuong{Not being referred in any Section/Paragraph.} \cuong{Use plural nouns for the title, such as ``Benchmark'', ``Failure Modes''.}\cuong{What are the purposes and mathematical formulation of metrics in the table?}\fi}
    \label{tab:rag_benchmark_comparison}
\end{table*}
This section reviews how recent studies benchmark RAG under failure modes. We first summarize these failure modes and then discuss how benchmark datasets and evaluation metrics are used to instantiate and compare them.

\noindent\textbf{Benchmarking Failure Modes.} Recent benchmark studies suggest that RAG failure is not one single problem, but a mix of attack susceptibility, retrieval brittleness, knowledge misalignment, and uncertainty under imperfect retrieval. On the adversarial side, \citet{liang2025saferag,zhang2025benchmarking,chen2025poisonarena} evaluate poisoning and injected evidence, showing that many retriever--generator pipelines remain vulnerable to realistic corpus manipulation. \citet{zhou2025emorag} further show that small query perturbations (e.g., symbolic or lexical edits) can shift retrieval enough to degrade final answers, while \citet{zheng2025knowshiftqa} examine knowledge conflicts between retrieved evidence and parametric memory. Beyond correctness, URAG~\citep{nguyen2026urag} benchmarks RAG uncertainty and shows that the typical relationship between higher accuracy and lower uncertainty can break under retrieval noise, resulting in confidently incorrect predictions. Taken together, these findings suggest that robustness benchmarks should assess not only adversarial resistance, but also retrieval stability, evidence-use behavior under conflict, and uncertainty calibration under noisy retrieval.

\noindent\textbf{Benchmark Datasets and Metrics.} We further examine the \emph{datasets} and \emph{evaluation metrics} used to benchmark these failure modes. \emph{(i) Datasets:} Many protocols instantiate the above stress settings on standard QA backbones. HotpotQA~\cite{yang2018hotpotqa}, Natural Questions~\cite{kwiatkowski-etal-2019-natural}, and SQuAD~\cite{rajpurkar-etal-2016-squad} are widely used to evaluate retrieval and generation robustness under controlled supervision. URAG~\citep{nguyen2026urag} further evaluates RAG reliability across diverse domains, including healthcare, programming, science, mathematics, and general text. \emph{(ii) Metrics:} Reported metrics generally evaluate \emph{answer quality}, \emph{retrieval quality}, and, more recently, \emph{uncertainty}. Answer quality is commonly measured using \emph{accuracy} and \emph{F1}, while adversarial benchmarks additionally report \emph{Attack Success Rate} (ASR)~\cite{liang2025saferag,zhang2025benchmarking,chen2025poisonarena}. Retrieval quality is typically quantified using \emph{precision}, \emph{recall}, and \emph{F1-score}. Complementarily, URAG measures uncertainty through conformal prediction using \emph{LAC} and \emph{APS} prediction-set sizes, enabling reliability to be evaluated jointly with answer accuracy.

\section{Variants of RAG}
\label{rag_application}
This section discusses emerging RAG structures, including \emph{GraphRAG}, \emph{Multimodal RAG} and \emph{Agentic RAG}, together with their safety challenges.

\noindent\textbf{GraphRAG.} Unlike standard RAG, which retrieves isolated text chunks, GraphRAG~\cite{graphrag2023} builds a graph over entities and relations in the corpus, summarizes it hierarchically, and retrieves structured summaries for generation. This graph-based retrieval design supports more global, corpus-level reasoning, but also introduces new safety challenges. Recent works examine the robustness of GraphRAG against adversarial attacks. \citet{Liang2025GraphRAGUF} show that the targeted \emph{corpus poisoning} attack that is effective for standard RAG becomes less effective under GraphRAG due to its graph indexing and distinct retrieval pipeline. Accordingly, later works develop corpus poisoning attacks tailored to GraphRAG by injecting entity or editing relational evidence to distort the induced knowledge graph~\cite{Liang2025GraphRAGUF,Zhao2025RAGSE,Wen2025AFW}. Beyond integrity attacks, GraphRAG also introduces privacy risks, the attacker can construct privacy extraction attacks tailored to GraphRAG to extract sensitive subgraphs or progressively reconstruct the underlying knowledge graph~\cite{Liu2025ExposingPR, Yang2026QueryEfficientAG}. 

\noindent\textbf{Multimodal RAG.} Multimodal RAG~\cite{chen-etal-2022-murag} extends conventional RAG by retrieving image-text pair evidence. This additional type of data is used mainly for visual grounding and reranking after retrieval. Compared with standard text-only RAG, this setting also introduces additional attack surfaces across both modalities. One line of work conducts \emph{corpus poisoning},~\citet{Ha2025MMPoisonRAGDM, Yu2025SpaVLMSP} show that injecting poisoned image-text pairs into the knowledge base can corrupt retrieval and mislead downstream generation, especially when the attacker can optimize both modalities under stronger white-box access. Complementarily, \citet{allawati2026multimodalragsystemsleak} study \emph{membership inference attacks}. They show that a carefully designed query paired with a target candidate image can induce the model to reveal whether that image is present in the database; once membership is established, the attacker may further extract the caption associated with the candidate image.

\noindent\textbf{Agentic and Tool-Augmented RAG.}
Conventional RAG is commonly modeled as a static three-component decomposition (corpus, retriever, generator). 
However, recent agentic and tool-augmented systems increasingly couple retrieval with persistent memory, 
external tools, planning loop, and interactions with open environments. These additional components substantially expand the attack surface: an adversarial artifact can not only 
affect the current retrieved context, but may also alter future memory retrieval, manipulate tool selection, 
inject malicious observations, or induce consequential downstream actions.

To capture the additional attack surfaces introduced by agentic and tool-augmented RAG, we extend the standard RAG formulation into a stateful agentic loop. Let $\mathcal{D}$ denote the document corpus, $M_t$ the persistent memory at step $t$, and $\mathcal{T}$ the available tool pool. An agentic RAG system consists of four modular functions. 
(1) A \emph{decision policy} $\pi:(x,y_t)\mapsto c_t$, where $c_t\in\{\textsc{Output},\textsc{Continue}\}$, determines whether the agent terminates or continues the interaction.
(2) A \emph{retriever} $R:(x,y_t;\mathcal{D},M_t,\mathcal{T})\mapsto z_t$ retrieves relevant documents, memories, or candidate tools.
(3) An \emph{LLM-based integration function} $F:(x,y_t,z_t,o_t)\mapsto(y_{t+1},a_t,m_t)$ integrates the retrieved resources with the latest tool- or environment-returned observation $o_t$ to produce the next agent state $y_{t+1}$, an action $a_t$, and a memory artifact $m_t$.
(4) A \emph{memory update function} $U:(M_t,m_t)\mapsto M_{t+1}$ incorporates the newly generated artifact into persistent memory.

\emph{Iterative Agentic RAG Process.} Given an input $x$, the process is initialized with $y_0=x$ and proceeds iteratively over $t=0,\ldots,T-1$. At each step, the system executes

\begin{equation}
    c_t = \pi(x,y_t),
\end{equation}

\begin{equation}
    z_t = R(x,y_t;\mathcal{D},M_t,\mathcal{T}),
\end{equation}

\begin{equation}
    (y_{t+1},a_t,m_t) = F(x,y_t,z_t,o_t),
\end{equation}

\begin{equation}
    M_{t+1} = U(M_t,m_t).
\end{equation}

Existing attacks expose vulnerabilities at several points in this loop. \textit{Memory and knowledge poisoning} targets resources that are repeatedly retrieved across interactions. 
AgentPoison \cite{chen2024agentpoison} poisons long-term memory or RAG knowledge bases so that 
triggered instructions retrieve malicious demonstrations while benign behavior is largely preserved. 
More importantly, direct write access is not always necessary: MINJA \cite{dong2025memory} shows that 
an attacker can inject malicious records into an agent's memory through query-only interaction, allowing 
the agent's own memory-update process to become an attack channel. Compared with conventional corpus poisoning, persistent-memory attacks are particularly concerning because 
a successful injection can influence later interactions rather than only the query under which the adversarial content was introduced.

A second attack surface arises from \textit{tool retrieval and selection}. Modern agents may first retrieve a small set of candidate tools from a large tool library before selecting 
one for execution. ToolHijacker \cite{shi2025promptinjectionattacktool} demonstrates that malicious tool descriptions can manipulate both of these stages, causing an agent to retrieve and select an attacker-controlled tool. This setting generalizes corpus poisoning from retrieving malicious \emph{evidence} to retrieving malicious 
\emph{capabilities}: the compromised resource may subsequently execute actions rather than merely alter 
the textual context.

A third class of attacks exploits \textit{untrusted observations returned by tools or environments}. 
InjecAgent \citep{zhan-etal-2024-injecagent} and AgentDojo \citep{debenedetti2024agentdojo} systematically 
demonstrate that indirect prompt injections embedded in external content can be returned through tool calls 
and interpreted by the agent as instructions, leading to privacy leakage or unauthorized actions. 
The same vulnerability appears in web agents, where malicious instructions embedded in webpage content or 
HTML accessibility representations can redirect agent behavior 
\citep{johnson-etal-2025-dangers}.

Finally, agentic systems amplify the \textit{consequences} of successful attacks. Whereas a compromised standard RAG system typically produces an incorrect or unsafe response, an agent may use the corrupted state to execute external actions. For example, backdoor attacks against embodied 
LLM-based decision-making systems can manipulate actions in autonomous-driving or robotic settings 
\citep{jiao2025trust}. Thus, the relevant security objective extends from preserving answer integrity to 
preserving the integrity of the entire retrieve--reason--act--update loop.

\section{Understanding RAG Robustness}
\label{app:Explainability}
\inlinecmt{\suhang{This section title might cause confusing, e.g., it sounds like RAG can provide explainable responses. Come up with a better section title, e.g., Understanding the Vulnerability of RAGs}}
Although it is well established that RAG systems are vulnerable to poisoned documents in the retrieval corpus, many questions about \textit{how and why these attacks succeed} remain open~\citep{cuong2024curious}. In this section, we review recent works that address these questions through explainability techniques, providing insights into the mechanisms that enable poisoned documents to influence RAG systems. 
\inlinecmt{\suhang{Some paragraphs below are quite long. Consider categorizing the works in each paragraph and adding some bullets, e.g., (i) xxx; (ii) xxx, to make them more structured}}

\noindent\textbf{Attribution Explanations.}
This line of work addresses the question, \textit{``Which context causes the output?''} and can be organized into three directions: \textit{(i) Document Attribution:} Identifies which retrieved or generated passages most influence the model’s output, helping trace whether unsafe behavior originates from external documents or generated context \cite{cohen2024contextcite}. \textit{(ii) Generated--Retrieved Conflict:} Goes beyond token-level attention by disentangling contributions from retrieved vs.\ generated context. \citet{tan2024blinded} show that models often favor generated context under conflict, even when retrieved evidence is correct, leading to potentially incorrect or unsafe outputs. \textit{(iii) Structural Evidence Bias:} Explains why certain contexts are more persuasive. \citet{chang2025externalknowledgepreferredllms} show that LLMs prefer coherent chains of evidence, which improves multi-hop reasoning but also makes well-structured misinformation more influential.

\noindent\textbf{Mechanistic Explanations.}
This line of work addresses the question, \textit{``How does the model resolve knowledge conflicts?''} and can be grouped into two directions: \textit{(i) Internal Conflict Signals:} \citet{zhao2024analysing} analyzes the residual stream and shows that signals for knowledge conflict and source selection emerge in intermediate layers, indicating whether the model is leaning toward parametric knowledge or contextual evidence before producing the final answer. \textit{(ii) Context Integration Bias:} \citet{tan2024blinded} shows that when retrieved evidence conflicts with generated context, models may favor the generated context even when the retrieved documents are correct. This explains why a poisoned or adversarial context overrides benign evidence.

\noindent\textbf{Structural Explanations.}
This line of work addresses the question, \textit{``Why is this evidence persuasive?''} and can be grouped into two directions: \textit{(i) Chain-of-Evidence Bias:} \citet{chang2025externalknowledgepreferredllms} shows that LLMs prefer evidence forming a coherent reasoning chain aligned with the query. While this can improve performance under noisy retrieval, it also makes well-structured but incorrect evidence more persuasive. \textit{(ii) Generated-Context Bias:} \citet{tan2024blinded} shows that generated contexts often appear more fluent and better aligned with the query than retrieved passages, causing models to overweight them even when retrieved evidence is correct.

\noindent\textbf{System-Level Explanations.}
This line of work addresses the question, \textit{``Why does safety degrade after adding retrieval?''} and can be grouped into two directions: \textit{(i) Safety Degradation Under Retrieval:} \citet{yu2025information} shows that adding retrieval can reduce refusal rates and amplify harmful or biased outputs, even when the retrieved information is itself accurate. \textit{(ii) Behavioral Regime Shift:} \citet{yu2025safety} explains the effect as a shift in the model’s decision: without retrieval, the model relies more on internal safety policies, whereas with retrieval, it prioritizes retrieving and synthesizing external evidence, which can override refusal behavior learned during alignment.

\section{Challenges and Future Directions}
\label{app:future_directions}
While attack and defense methods have reviewed serious vulnerabilities in RAG and defense mechanisms, several critical challenges remain. Based on the discussions above, we highlight primary directions for future research.

\noindent\textbf{Lack of Transferability.} Several RAG attacks assume strong white-box access to the retriever or generator, enabling direct optimization over embeddings, gradients, or token-level probabilities~\cite{wang2025jointgcgunifiedgradientbasedpoisoning}. However, these assumptions do not always align with realistic deployment settings, where attackers often face only black-box or limited-query access. As a result, attack effectiveness may degrade substantially when the same method is adapted across different models~\cite{chaudhari2024phantom}.

\noindent\textbf{Lack of Stealthiness.} Although many poisoning attacks are effective at manipulating retrieval and generation, their injected documents or optimized sequences are often not sufficiently stealthy~\cite{poisonedrag2024}. In particular, some methods produce adversarial texts that sacrifice semantic coherence or naturalness in exchange for stronger retrievability or attack success. For instance, \citet{su2025corpus} note that the generated adversarial sequences can be semantically unnatural, making the poisoned corpus easier to detect through human inspection or filtering-based defenses.

\noindent\textbf{Adversarial Attacks on Emerging RAG.} As RAG has expanded into emerging paradigms such as \emph{GraphRAG} and \emph{multimodal RAG}, the study of adversarial attacks in these settings remains limited. In particular, \emph{privacy attacks} on graph-based and multimodal retrieval systems represent a promising direction for future research, as attackers can exploit these architectures to extract structure and visible data which violate property risks dangerously.

\noindent\textbf{Traceback Defenses Remain Largely Diagnostic rather than Preventive.} Attribution accuracy can degrade under highly entangled contexts, paraphrased poisoning, or colluding benign and malicious documents. Many methods rely on heuristic influence measures or model-internal signals that lack formal guarantees, while provable approaches such as watermarking assume control over dataset construction. Scalability to very long contexts, multimodal settings, and agentic workflows remains an open challenge, as does standardizing evaluation metrics for attribution faithfulness and robustness. Bridging traceback with automated remediation and real-time defense integration is a promising direction for future research.

\noindent\textbf{Diverse-Query Attacks.} Existing poisoning attacks exhibit a trade-off between target-specific control and broad query coverage. Query-specific methods can achieve strong manipulation for designated queries, but require repeated optimization and may become less effective as additional clean evidence is retrieved. In contrast, universal attacks improve scalability by optimizing a shared set of adversarial texts over representative query clusters, but their effectiveness depends on how well the sampled queries and clustering structure capture the underlying query distribution. Future work should evaluate scalability across query distributions, retrieval settings, corpus updates, and agentic workflows.

\noindent\textbf{Fine-Grained Source Attribution.} Privacy leakage in RAG may arise from multiple sources, including retrieved private documents, memorized knowledge in the generator, semantically related evidence, or correlations with publicly available data. As a result, observing sensitive information in the model output does not necessarily reveal which component is responsible for the leakage. Future benchmarks should disentangle these sources through controlled corpus membership, retrieval ablations, paraphrased variants, and source-level attribution.

\definecolor{privacyLight}{HTML}{FFF9E8}   
\definecolor{accuracyLight}{HTML}{EEF9F1}  
\definecolor{fairnessLight}{HTML}{F2F8FF}  
\definecolor{headerBg}{HTML}{2C3E50}       
\definecolor{headerFg}{HTML}{FFFFFF}       
\definecolor{subrowA}{HTML}{F7FBFF}        
\definecolor{subrowB}{HTML}{FFFDF5}        

\begin{table*}[tb!]
\centering
\renewcommand{\arraystretch}{1.6}
\setlength{\tabcolsep}{8pt}
\resizebox{\textwidth}{!}{%
\large
\setlength{\aboverulesep}{0pt}
\setlength{\belowrulesep}{0pt}
\begin{tabular}{llll}
\toprule

\cellcolor{headerBg}\textcolor{white}{\textit{Objectives}} &
\cellcolor{headerBg}\textcolor{white}{\textit{Targeting}} &
\cellcolor{headerBg}\textcolor{white}{\textit{Methods}} &
\cellcolor{headerBg}\textcolor{white}{\textit{References}} \\
\midrule

\multirow{10}{*}{\textbf{Accuracy}} &
\multirow{10}{*}{Targeted} &
 &
\cellcolor{accuracyLight}\citet{paradoxRAG2025}, \citet{xi2025riprag}, \citet{chang2025one},
  \citet{zhuang2024vec2text}, \citet{10.5555/3766078.3766274}, \\
 & & &
\cellcolor{accuracyLight}\citet{li2025cparagcovertpoisoningattacksretrievalaugmented}, \citet{cho-etal-2024-typos}, \citet{song-etal-2025-silent},
  \citet{wu2025admitfewshotknowledgepoisoning}, \\
 & & Corpus Poisoning &
\cellcolor{accuracyLight}\citet{Chen2024BlackBoxOM}, \citet{10.1007/978-3-031-88717-8_18},
  \citet{poisonedrag2024}, \citet{su2025corpus},
  \citet{shao2025poisoncraftpracticalpoisoningretrievalaugmented}, \\
 & & &
\cellcolor{accuracyLight}\citet{chen2025eyesonmescalableragpoisoning},
  \citet{li2025tokenlevelpreciseattackrag},
  \citet{wang2025jointgcgunifiedgradientbasedpoisoning},
  \citet{suo2025hoistpetardinducingguardrails}, \\
 & & &
\cellcolor{accuracyLight}\citet{xian2025vulnerabilityrag}, \citet{zhang2024adversarial} \\
\cmidrule{3-4}

 & &
\multirow{2}{*}{Backdoor Attack} &
\cellcolor{accuracyLight}\citet{chaudhari2024phantom}, \citet{chaturvedi-etal-2025-aip},
  \citet{jiao2024prattack}, \citet{cheng2024trojanrag}, \\
 & & &
\cellcolor{accuracyLight}\citet{jiao2025trust}, \citet{chen2024agentpoison},
  \citet{patlan2025realaiagentsfake} \\
\cmidrule{3-4}

 & &
Prompt Injection &
\cellcolor{accuracyLight}\citet{neuroexecpasquini} \\
\cmidrule{2-4}

 &
Untargeted &
Corpus Poisoning &
\cellcolor{accuracyLight}\citet{tamber-lin-2025-illusions}, \citet{tan-etal-2024-glue},
  \citet{geng2025unic}, \citet{wang2025tricking} \\
\midrule
\multirow{2}{*}{\textbf{Fairness}} &
\multirow{2}{*}{Targeted} &
Corpus Poisoning &
\cellcolor{fairnessLight}\citet{wang2025biasrag} \\
\cmidrule{3-4}
 & &
Backdoor Attack &
\cellcolor{fairnessLight}\citet{bagwe2025ragunfair} \\
\midrule

\multirow{5}{*}{\textbf{Privacy}} &
\multirow{3}{*}{Targeted} &
\multirow{1}{*}{Membership Inference} &
\cellcolor{privacyLight}\citet{10.1145/3696410.3714771}, \citet{mia2025}, \citet{GeneratingIsBelieving}, \citet{wang2025ragleaks}. \\
\cmidrule{3-4}
 & &
Privacy Extraction &
\cellcolor{privacyLight}\citet{Jiang2024FeedbackGuidedEO}, \citet{qi2025follow}, \citet{zeng-etal-2024-good}, \citet{chen2025finegrainedprivacyextractionretrievalaugmented}, \citet{peng2025dataextractionattacksretrievalaugmented} \\
\cmidrule{3-4}
 & &
Confused-Deputy &
\cellcolor{privacyLight}\citet{roychowdhury2024confusedpilotconfuseddeputyrisks} \\
\cmidrule{2-4}
& Untargeted & Privacy Extraction &
\cellcolor{privacyLight}\citet{zeng-etal-2024-good}, \citet{Jiang2024FeedbackGuidedEO} \\
\bottomrule
\end{tabular}%
}
\caption{%
  \textbf{Taxonomy of attacks in retrieval-augmented generation (RAG).}
  Attacks are categorized by objective (\emph{accuracy}, \emph{fairness}, \emph{privacy}),
  targeting strategy (\emph{targeted} vs.\ \emph{untargeted}), and attack method.
}
\label{tab:rag_attack_taxonomy}
\end{table*}

\definecolor{retrievalLight}{HTML}{F6EFFB}   
\definecolor{rerankLight}{HTML}{F2F8FF}      
\definecolor{tracebackLight}{HTML}{FFF9E8}   
\definecolor{headerBg}{HTML}{2C3E50}         
\definecolor{subrowA}{HTML}{F7FBFF}          
\definecolor{subrowB}{HTML}{FFFDF5}          
\definecolor{subrowC}{HTML}{F5F0FA}          
\definecolor{generationLight}{HTML}{EEF9F1}  
\begin{table*}[tb!]
\centering
\renewcommand{\arraystretch}{1.6}
\setlength{\tabcolsep}{8pt}
\resizebox{\textwidth}{!}{%
\large
\setlength{\aboverulesep}{0pt}
\setlength{\belowrulesep}{0pt}
\begin{tabular}{llll}
\toprule

\cellcolor{headerBg}\textcolor{white}{\textit{Stages}} &
\cellcolor{headerBg}\textcolor{white}{\textit{Categories}} &
\cellcolor{headerBg}\textcolor{white}{\textit{Methods}} &
\cellcolor{headerBg}\textcolor{white}{\textit{References}} \\
\midrule

\multirow{6}{*}{\textbf{Retrieval}} &
\multirow{4}{*}{Robust Retrieval} &
Diagnostic Resources &
\cellcolor{subrowC}\citet{fayyaz2025collapsedenseretrieversshort},
  \citet{usmb_semantic_similarity} \\
\cmidrule{3-4}
 & &
Evidence-Centric Retrieval &
\cellcolor{subrowC}\citet{chang2025externalknowledgepreferredllms} \\
\cmidrule{3-4}
 & &
\multirow{2}{*}{Corpus Sanitization} &
\cellcolor{subrowC}\citet{jia2025ragrankusingpagerankcounter},
  \citet{si2025seconragtwostagesemanticfiltering}, \\
 & & &
\cellcolor{subrowC}\citet{11075247}, \citet{10885343},
  \citet{mochizuki2026exposingcitationvulnerabilitiesgenerative} \\
\cmidrule{2-4}

 &
\multirow{2}{*}{\shortstack[l]{Robust Training\\for Retriever}} &
Hard Negative Training &
\cellcolor{subrowA}\citet{qu-etal-2021-rocketqa},
  \citet{thakur2025hardnegativeshardlessons} \\
\cmidrule{3-4}
 & &
Perturbation Training &
\cellcolor{subrowA}\citet{tasawong2023typorobustrepresentationlearningdense} \\
\midrule

\textbf{Rerank} &
Graph-Based Reranking &
Similarity Graph &
\cellcolor{rerankLight}\citet{grada} \\
\midrule

\multirow{10}{*}{\textbf{Generation}} &
\multirow{6}{*}{\shortstack[l]{Proactive\\Defenses}} &
\multirow{3}{*}{Filtering} &
\cellcolor{generationLight}\citet{zhou2025trustrag}, \citet{kim2025rescuingunpoisonedefficientdefense},
  \citet{cheng2025secureretrievalaugmentedgenerationpoisoning},
  \citet{shen2026reliabilityrag}, \\
 & & &
\cellcolor{generationLight}\citet{choudhary2025stealthlensrethinkingattacks},
  \citet{kim-etal-2025-safeguarding}, \\
 & & &
\cellcolor{generationLight}\citet{hwang-etal-2025-retrieval},
  \citet{shen2026reliabilityrag} \\

\cmidrule{3-4}
 & &
\multirow{2}{*}{Detection} &
\cellcolor{generationLight}\citet{tan-etal-2025-revprag},
  \citet{zou2025pishielddetectingpromptinjection}, \citet{10992559}, \\
 & & &
\cellcolor{generationLight}\citet{hong-etal-2024-gullible},
  \citet{shi2025promptarmorsimpleeffectiveprompt}, \citet{alon2024detecting} \\
\cmidrule{2-4}

 &
\multirow{5}{*}{\shortstack[l]{Passive\\Defenses}} &
\multirow{3}{*}{\shortstack[l]{Robust Reasoning\\and Generation}} &
\cellcolor{subrowB}\citet{xiang2024certifiably}, \citet{weiretrieval}, \\
 & & &
\cellcolor{subrowB}\citet{wang-etal-2025-astute}, \citet{dai2025after},
  \citet{chen2025rethinking}, \citet{xiong2025icr}, \\
 & & &
\cellcolor{subrowB}\citet{shi2025making}, \citet{yao2025controlnet},
  \citet{chen2025scalable} \\

\cmidrule{3-4}
 & &
\multirow{2}{*}{\shortstack[l]{Robust Training\\for Generator}} &
\cellcolor{subrowB}\citet{yoran2024making}, \citet{fang-etal-2024-enhancing},
  \citet{10.1145/3726302.3730078}, \\
 & & &
\cellcolor{subrowB}\citet{yan-etal-2025-rpo}, \citet{huang2025to},
  \citet{wei2025instructrag} \\
\midrule

\multirow{3}{*}{\textbf{Traceback}} &
\multirow{3}{*}{Attribution} &
Document Level &
\cellcolor{tracebackLight}\citet{zhang2025taught}, \citet{zhang2025traceback} \\
\cmidrule{3-4}
 & &
Context \& Token Level &
\cellcolor{tracebackLight}\citet{cohen2024contextcite}, \citet{wang2025tracllm},
  \citet{wang2025attntrace} \\
\cmidrule{3-4}
 & &
Dataset Level &
\cellcolor{tracebackLight}\citet{jovanovic2024ward} \\
\bottomrule
\end{tabular}%
}
\caption{%
  \textbf{Taxonomy of defenses for retrieval-augmented generation (RAG) robustness.}
  Defenses are organized by their position in the RAG pipeline.
}
\label{tab:rag_defense_taxonomy}
\end{table*}
\definecolor{privacyLight}{HTML}{DDEEFF}   
\definecolor{accuracyLight}{HTML}{FFF3CD}  
\definecolor{fairnessLight}{HTML}{D6F5E3}  
\definecolor{headerBg}{HTML}{2C3E50}       
\definecolor{headerFg}{HTML}{FFFFFF}       
\definecolor{subrowA}{HTML}{F7FBFF}        
\definecolor{subrowB}{HTML}{FFFDF5}        
\definecolor{retrievalLight}{HTML}{F6EFFB}   
\definecolor{whiteBoxLight}{HTML}{FDEEEF}  
\begin{table*}[t]
\centering
\small
\renewcommand{\arraystretch}{1.2}
\setlength{\tabcolsep}{8pt}
\setlength{\aboverulesep}{0pt}
\setlength{\belowrulesep}{0pt}
\begin{tabular}{
>{\centering\arraybackslash}m{2.25cm}
>{\raggedright\arraybackslash}m{3cm}
>{\raggedright\arraybackslash}m{9.1cm}
}
\toprule
\cellcolor{headerBg}\textcolor{headerFg}{\textit{Stages}} &
\cellcolor{headerBg}\textcolor{headerFg}{\textit{Techniques}} &
\cellcolor{headerBg}\textcolor{headerFg}{\textit{Description}} \\
\midrule
\multirow{2}{*}{\textbf{Retrieval}}
& \cellcolor{subrowA}Robust Retrieval
& \cellcolor{subrowA}Improve retriever through test-time mechanisms (e.g., CoE retrieval, corpus sanitization). \\
& \cellcolor{subrowA}Robust Training
& \cellcolor{subrowA}Train the retriever with adversarial or noisy data to improve robustness. \\
\midrule
{\textbf{Rerank}}
& \cellcolor{subrowB}Consistency Reranking
& \cellcolor{subrowB}Reorder retrieved candidates to promote benign and relevant documents while pushing suspicious documents below the final top-$k$ cutoff. \\
\midrule
\multirow{6}{*}{\textbf{Generation}}
& \cellcolor{retrievalLight}Filtering
& \cellcolor{retrievalLight}Remove noisy, poisoned, or prompt-injected documents from the retrieved context before generation through consistency, influence, mechanistic and reliability-aware filtering. \\
& \cellcolor{retrievalLight}Detection
& \cellcolor{retrievalLight}Identify and flag suspicious retrieved document or abnormal generation signals without necessarily removing them. \\
& \cellcolor{retrievalLight}Robust Reasoning
& \cellcolor{retrievalLight}Improve reasoning through test-time mechanisms (e.g., alignment, layer guided attention). \\
& \cellcolor{retrievalLight}Robust Training
& \cellcolor{retrievalLight}Train the generator with adversarial, irrelevant, noisy or counterfactual contexts to internalize defensive behavior. \\
\midrule
{\textbf{Traceback}}
& \cellcolor{whiteBoxLight}Traceback / Attribution
& \cellcolor{whiteBoxLight}Trace the generated output back to the responsible document, passage, or data source after a harmful response is observed. \\
\bottomrule
\end{tabular}
\caption{\textbf{Defense taxonomy for RAG robustness.} We summarize defense techniques by stage, together with their descriptions.}
\label{tab:rag_defense_background}
\end{table*}

\end{document}